\documentclass[fleqn,usenatbib,useAMS]{mnras}

\usepackage{graphicx}	
\usepackage{amsmath}	
\usepackage{amssymb}	
\usepackage{multicol}        
\usepackage{bm}		
\usepackage{pdflscape}	
\usepackage{subfig}
\usepackage{multirow}

\usepackage{scalerel,tikz}
\usetikzlibrary{svg.path}
\definecolor{orcidlogocol}{HTML}{A6CE39}
\tikzset{orcidlogo/.pic={
 \fill[orcidlogocol] svg{M256,128c0,70.7-57.3,128-128,128C57.3,256,0,198.7,0,128C0,57.3,57.3,0,128,0C198.7,0,256,57.3,256,128z};
 \fill[white] svg{M86.3,186.2H70.9V79.1h15.4v48.4V186.2z}
 svg{M108.9,79.1h41.6c39.6,0,57,28.3,57,53.6c0,27.5-21.5,53.6-56.8,53.6h-41.8V79.1z M124.3,172.4h24.5c34.9,0,42.9-26.5,42.9-39.7c0-21.5-13.7-39.7-43.7-39.7h-23.7V172.4z}
 svg{M88.7,56.8c0,5.5-4.5,10.1-10.1,10.1c-5.6,0-10.1-4.6-10.1-10.1c0-5.6,4.5-10.1,10.1-10.1C84.2,46.7,88.7,51.3,88.7,56.8z};
}}
\newcommand\orcidicon[1]{\href{https://orcid.org/#1}{\mbox{\scalerel*{
\begin{tikzpicture}[yscale=-1,transform shape]
\pic{orcidlogo};
\end{tikzpicture}
}{|}}}}

\usepackage[T1]{fontenc}
\usepackage{ae,aecompl}

\usepackage{newtxtext,newtxmath}

\title[Stellar halo striations]{Stellar halo striations from assumptions of axisymmetry}

\author[E. Y. Davies et al.]{
Elliot Y. Davies~\orcidicon{0000-0001-5996-4072}$^{1}$\thanks{E-mail: eyd20@cam.ac.uk},
Adam M. Dillamore~\orcidicon{0000-0003-0807-5261}$^{1}$,
Vasily Belokurov~\orcidicon{0000-0002-0038-9584}$^{1}$
and N. Wyn Evans~\orcidicon{0000-0002-5981-7360}$^{1}$.
\\
$^{1}$Institute of Astronomy, University of Cambridge, Madingley Road, Cambridge CB3 0HA, UK}

\date{Last updated 2020 June 10; in original form 2013 September 5}

\pubyear{2023}

\begin{document}
\label{firstpage}
\pagerange{\pageref{firstpage}--\pageref{lastpage}}
\maketitle

\begin{abstract}
Motivated by the LMC's impact on the integral of motion space of the stellar halo, we run an $N$-body merger simulation to produce a population of halo-like stars. We subsequently move to a test particle simulation, in which the LMC perturbs this debris. When an axisymmetric potential is assumed for the final snapshot of the $N$-body merger remnant, a series of vertical striations in $(L_z, E)$ space form as the LMC approaches its pericentre. These result from the formation of overdensities in angular momentum owing to a relationship between the precession rate of near radial orbits and the torquing of these orbits by the LMC. This effect is heavily dependent on the shape of the inner potential. If a quadrupole component of the potential is included these striations become significantly less apparent due to the difference in precession rate between the two potentials. The absence of these features in data, and the dramatic change in orbital plane precession rate, discourages the use of an axisymmetric potential for highly eccentric orbits accreted from a massive GSE-like merger. Given the link between appearance of these striations and the shape of the potential, this effect may provide a new method of constraining the axisymmetry of the halo.
\end{abstract}




\begingroup
\let\clearpage\relax
\endgroup
\newpage

\section{Introduction}

The stellar halo of the Milky Way (MW) is rich with kinematic and chemical information that provides us a window into the history of our home Galaxy. Like any typical galaxy, the MW likely built up its dark matter content by the accumulated merging of smaller galaxies over time, in accordance with the cold dark matter model of cosmology \citep[][]{white1991galaxy}. The stellar halo of our Galaxy is dominated by content from its most recent major merger, dubbed the \textit{Gaia} Sausage / Enceladus (GSE). The GSE is an ancient $(1 < z < 2)$ merger \citep[][]{belokurov2018coformation, helmi2018merger}, estimated to contribute as much as 2/3 of the stellar halo \citep[][]{fattahi2019origin, dillamore2022merger,naidu2021reconstructing}. The discovery of distinct halo-like stars, with high radial velocity anisotropy \citep[e.g.][]{lancaster2019halos, necib2019inferred, bird2021constraints, iorio2021chemo}, led many to the conclusion that the GSE deposited stars onto the early MW with a highly eccentric orbit. Furthermore, GSE-like merger simulations with high mass ratio $(q \gtrsim 0.1)$ and low initial circularity $(\eta \lesssim 0.5)$, show that such satellites often sink deep within their host and undergo radializaton \citep[e.g.][]{vasiliev2022radialization}. Rather uniquely, the merger history of the MW following the GSE is thought to be much more quiet  \citep[e.g.][]{deason2013broken,naidu2020evidence,Evans2020}. Given this reasonably distinct formation history, and the predicted nature of the GSE's infall trajectory, one would anticipate that the shape of the inner halo would also be heavily influenced by the infall of the its most recent major merger. However, most work until the recent past assumed an oblate spheroidal shape for the stellar halo of the MW, as presented by \citet[][]{blandhawthorn2016galaxy}. However, this assumption of axisymmetry has been more seriously challenged in recent years by \citet[][]{iorio2019shape,han2022stellar, han2022tilt} amongst others. Moreover, a wide variety of measured radial density profiles of the stellar halo is a further indicator that an axisymmetric shape may be not truly representative \citep[e.g.][]{deason2011milky, lowing2015creating, hernitschek2018profile}. The current consensus is that the density of the stellar halo, within 30 kpc of the Galactic centre, has been shown to have axis ratios of about $10:8:4.5-7$ \citep[][]{iorio2019shape, naidu2021reconstructing}, where the major axis is at a $20-30$ degree angle to the plane. Additionally, the resulting apocentric pile-up of the GSE merger approximately coincides with the positions of the Hercules-Aquila Cloud and the Virgo Overdensitiy \citep[][]{deason2018apocenter, simion2018spectroscopic, simion2019common}. Assumptions of a spheroidal dark halo potential have led to a number of interesting conclusions, including the possibility that the GSE may not be as old as we thought, given the time at which the HAC and VOD should have phase-mixed away \citep[][]{donlon2019virgo, donlon2020milky, donlon2022local}.

In spherically symmetric potentials, angular momentum is conserved. However, in flattened axisymmetric potentials the angular momentum vector of an orbit will precess around the symmetry axis, and thus orbits will continually evolve. As discussed by \citet[][]{ibata2001great, erkal2016stray} in the context of stellar streams, the orbital plane precession rate of an arbitrary orbit is linked directly to the amount of flattening (i.e. the shape) of an axisymmetric potential. Of course a plethora of work, pre-dating studies of stellar streams, exists on the topic of orbital plane precession \citep[e.g.][]{steinmancameron1990simple, murray1999solar}. While the angular momentum vector aligned with the symmetry axis is conserved for an axisymmetric potential, this is not the case for a triaxial potential, as there is no rotational symmetry. Evidently, the orbital plane precession rate in a triaxial potential is far more complex, but is still linked to the exact shape of the potential. Therefore, just as with stellar streams, one expects that the precession rate of eccentric GSE-like orbits could illuminate the shape of the inner potential. Work by \citet[][]{dodge2023dynamics} into the tilting of the stellar disk shows how easily orbital plane precession rates can be affected by a GSE-like merger. In this report, we present how the orbital plane of highly eccentric accreted halo-like orbits precesses in different potentials. Given the uncertainty in the exact shape in the inner GSE debris, and the sensitivity of highly radial orbits to the centre of the potential, any difference is crucial to understand.

Aside from the GSE, the MW is surrounded by many smaller galaxies and globular clusters \citep[e.g.][]{pace2022proper}, that are themselves evidence of the past and ongoing formation of the Galaxy. The largest of the MW's satellites is the Large Magellanic Cloud (LMC), likely making its first pericentric passage of $r_{\rm peri}\sim 50$ kpc about the MW. With a mass of $M_{\rm LMC} \simeq 1.4\times10^{11}$ M$_{\odot}$ \citep[][]{erkal2019total}, which is a significant fraction of the MW's mass contained within $r_{\rm peri}$, the influence that the LMC will have on the Galactic halo is significant. Evidence of the LMC's effect on the stellar halo has already been provided; studies have shown its impact both on stellar streams within the halo \citep[e.g.][]{erkal2019total, vasiliev2021tango, shipp2021measuring, dillamore2022impact}, and on the shape of the halo on large scales~\cite[e.g.][]{garavito2019hunting, belokurov2019pisces, petersen2020reflex, conroy2021allsky}. Interestingly, \citet[][]{iorio2019shape} noted that the major axis of the GSE debris points roughly along the trajectory of the LMC, which has brought speculation of shared origins or other connections between the two. 

The standard model of galaxy formation predicts that the MWs stellar halo is populated with `substructure' from the remnants of its past mergers, either in the form of stellar streams \citep[e.g.][]{belokurov2006field} or moving clumps of stars with a shared origin \citep[][]{helmi1999building}. In the search for substructure in the stellar halo many have turned to `integral of motion space', often energy and angular momentum space, as an alternative to simple searches in configuration space \citep[e.g.][]{helmi2017box, myeong2018discovery, dodd2023gaia}. Additionally, substructure within the phase space of our Galaxy offers further insight into its formation history \citep[e.g.][]{antoja2018dynamically, belokurov2023energy}. In the era of the \textit{Gaia} mission \citep[][]{Gaia}, the search for Galactic substructure is more extensive than ever, and now numerous sub-populations of the stellar halo have been identified and labelled thanks to the over 25 million stars with full 6-d information. 

However, the violent formation history of any galaxy could result in changes to its potential, such that these substructures detected in integral of motion or phase space space may be misleading. The Milky Way is no exception, and it has already been shown how large bodies can mimic the creation of smaller substructure \citep[e.g.][]{jeanbabtiste2017kinematic, belokurov2023energy} or perturb existing substructure in potentially deceptive ways \citep[e.g.][]{erkal2019total, vasiliev2021tango}. Due to its mass and close trajectory, the LMC is likely to be one of the major culprits responsible for disturbing the integral of motion space of the Milky Way, specifically the outer halo. As we now begin to probe regions of the outer halo that extend beyond the LMCs pericentre \citep[e.g.][]{chandra2022distant}, it is vital to understand any effect our largest local neighbour may induce.

Motivated by the influence that the LMC may have on the integral of motion space of the stellar halo, we ran an $N$-body simulation to recreate GSE-like orbits. After a few Gyr of phase mixing, we moved to a static host potential and subsequently integrated these GSE-like orbits in a potential that consisted of the static host and an perturbing LMC. As the LMC approached pericentre, we noticed the appearance of vertical striations in $(L_z, E)$. Given the absence of these features in data, we examined further and found that an assumption of axisymmetric in our static host was responsible. The structure of this report is as follows. In Sec.~\ref{sec:method} we briefly detail the method of our simulations used throughout this work. Sec.~\ref{sec:mergerdebris} presents, analyses and explains the formation of $(L_z, E)$ striations in simulated GSE-like merger debris, assuming an axisymmetric potential. We follow this up with a comparison with a quadrupole potential in Sec.~\ref{sec:compare}. Lastly, we sum up in Sec.~\ref{sec:summary}.

\section{Simulation Method}\label{sec:method}

Many of the simulations utilised in this work have already been described in detail elsewhere \citep{davies2023ironing}. We repeat the essential details and outline any differences.

\begin{figure}
    \centering
    \includegraphics[width=0.8\columnwidth]{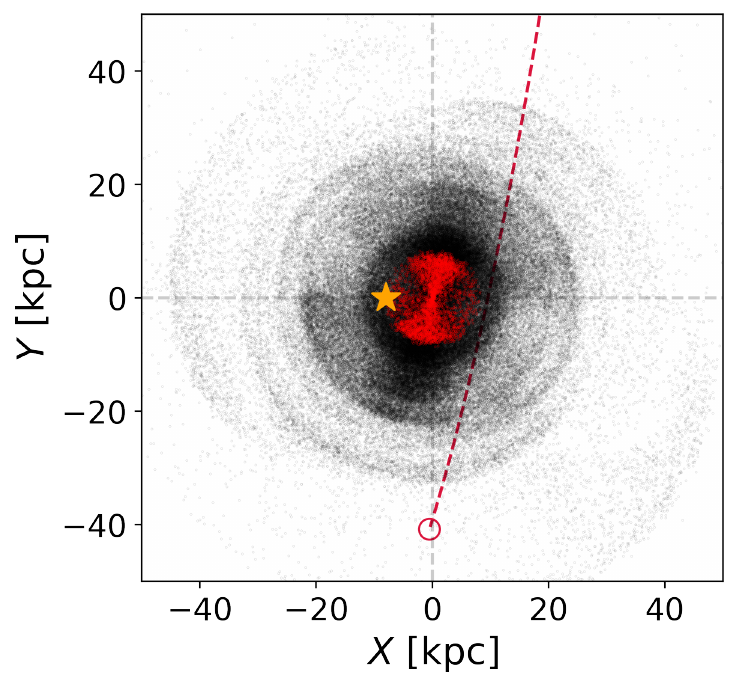}
    \caption{Top down view of the merged satellite debris, at the end of the $N$-body simulation, before the LMC interaction. The black particles show the entire population, whereas the red particles show the selected low energy debris. The red dashed line shows the trajectory of the LMC, with the open red circle showing its final position at the end of the simulation. The orange star shows the present position of the sun.}
    \label{fig:debris}
\end{figure}

\subsection{GSE-like merger}

\begin{table}
\centering
\caption{Structural parameters of the host in the initial $N$-body merger simulation described in Sec.~\ref{sec:method}}
\label{tab:host_params}
\resizebox{\columnwidth}{!}{%
\begin{tabular}{|l|l|l|l|}
\hline
Component & Type & Parameter & Value \\ \hline\hline
\multirow{2}{*}{Halo} & \multirow{2}{*}{Hernquist} & total mass {[}$10^{11}$ M$_{\odot}${]} & 5 \\
 &  & scale radius {[}kpc{]} & 15 \\ \hline
\multirow{3}{*}{Disk} & \multirow{3}{*}{Exponential} & total mass {[}$10^{9}$ M$_{\odot}${]} & 5 \\
 &  & scale radius {[}kpc{]} & 2 \\
 &  & scale height {[}kpc{]} & 0.5 \\ \hline
\multirow{3}{*}{Bulge} & \multirow{3}{*}{S\'{e}rsic} & total mass {[}$10^{9}$ M$_{\odot}${]} & 2.25 \\
 &  & scale radius & 0.8 \\
 &  & S\'{e}rsic index & 2.0 \\ \hline
\end{tabular}%
}
\end{table}

\begin{table}
\centering
\caption{Structural parameters of the satellite in the initial $N$-body merger simulation described in Sec.~\ref{sec:method}}
\label{tab:sat_params}
\resizebox{\columnwidth}{!}{%
\begin{tabular}{|l|l|l|l|}
\hline
Component & Type & Parameter & Value \\ \hline\hline
\multirow{2}{*}{Halo} & \multirow{2}{*}{Hernquist} & total mass {[}$10^{11}$ M$_{\odot}${]} & 2 \\
 &  & scale radius {[}kpc{]} & 12 \\ \hline
\multirow{3}{*}{Disk} & \multirow{3}{*}{Exponential} & total mass {[}$10^{9}$ M$_{\odot}${]} & 2.25 \\
 &  & scale radius {[}kpc{]} & 1.7 \\
 &  & scale height {[}kpc{]} & 1.0 \\ \hline
\end{tabular}%
}
\end{table}

To simulate the GSE debris of the stellar halo, we run an $N$-body merger simulation very similar to the one described in §4.1 of \citet[][]{davies2023ironing}. The MW-like host is comprised of a combined disk, bulge and halo potential, whereas the GSE-like satellite is just made of a disk and a halo component. In both cases, the halo component is described by a Hernquist potential \citep[][]{hernquist1990analytical}, the disk component by an exponential profile, and the bulge component by a S\'{e}rsic profile \citep[][]{influence1963sersic}. The $N$-body is ran for a total of 5 Gyr before any additional potential components are introduced, allowing the satellite to phase mix in isolation. The satellite galaxy is represented by $2 \times 10^5$ stellar particles and $8\times10^5$ dark particles, and has a total mass of $2\times10^{11}$ M$_{\odot}$. The host galaxy is represented by twice as many particles, and has a total mass of $5\times10^{11}$ M$_{\odot}$. Therefore, the mass of the merger remnant, enclosed within $100$ kpc, is equal to about $4/5$ that of the \textsc{MilkyWayPotential} potential in \textsc{Gala} \citep[][]{price-whelan2017gala}. The shape of the merger remnant is weakly triaxial, with axis ratios 1 : 0.88 : 0.38. We summarise the structural parameters of the MW-like host in Table.~\ref{tab:host_params} and the GSE-like satellite in Table.~\ref{tab:sat_params}. The satellite is placed on a prograde orbit with an inclination of 30$^{\circ}$ and a circularity of $\eta = 0.5$, where $\eta = L/L_{\rm circ}(E)$ is the ratio of total angular momentum to the angular momentum of a circular orbit of the same energy energy $E$.

After evolving the simulation for 5 Gyr with the \textsc{gyrfalcON} code \citep[][]{dehnen2002falcon}, we create a static axisymmetric potential from the final snapshot represented by a multipole expansion, described in detail in Section 2 of \citep[][]{vasiliev2019agama}. We save the final positions and velocities of the satellite’s stellar particles, and apply a rotation matrix such that the long axis of the debris is at a 70 degree orientation to the $x$-axis, roughly in line with the major axis of the GSE debris \citep[][]{iorio2019shape, naidu2021reconstructing}. This ensures an appropriate orientation of the debris relative to the LMC's trajectory. Lastly, we integrate the orbits of just the stellar satellite particles in the static potential of the (host plus satellite) merger remnant, using the \textsc{Agama} code \citep[][]{vasiliev2019agama}. In Fig.~\ref{fig:debris}, we provide a visual representation of the debris after the 5 Gyr $N$-body simulation. For a comparison with the static potentials, we also evolve the $N$-body simulation for a further 3 Gyr, for a combined total of 8 Gyr $N$-body. This combined time of 8 Gyr for the entire simulation is at the lower limit of time age of the GSE merger.

\subsection{Static potentials}

\begin{table}
\centering
\caption{The $l=2$ multipole power $C_l$ for the axisymmetric and quadrupole potential, calculated at the solar radius ($\sim 8$ kpc) using Eq.~\ref{eq:power_coeff}. The columns detailing the power are normalised by the axisymmetric $l=0$ power $C_0^{\rm (axi)}$.}
\label{tab:coefficients}
\resizebox{\columnwidth}{!}{%
\begin{tabular}{|l|l|l|l|}
\hline
Potential ($m_{\rm max}$) & $C_2/C_0^{\rm (axi)}$ & $C_4/C_0^{\rm (axi)}$ & $C_6/C_0^{\rm (axi)}$ \\ \hline \hline
axisymmetric (0) & $9.16\times10^{-4}$ & $1.77\times10^{-5}$ & $6.72\times10^{-7}$\\ \hline
quadrupole (2) & $1.22\times10^{-3}$ & $2.26\times10^{-5}$ & $8.44\times10^{-7}$ \\ \hline
\end{tabular}%
}
\end{table}

As described above, we integrate the orbits of $N$-body debris as test particles in a static potential, generated from entire merger debris using the multipole expansion tool in \textsc{Agama} \citep[][]{vasiliev2019agama}. In this approach, the potential is represented by a sum of spherical-harmonics: 

\begin{align}
\begin{split}
    \Phi(r,\theta,\phi) &= \sum_{l=0}^{l=l_{\rm max}}\sum_{m = -m_0(l)}^{m = m_0(l)}\Phi_{l,m}(r)\sqrt{4\pi}\,\Tilde{P}^m_l(\cos\theta) \, {\rm trig} \, m\phi \\
    {\rm trig} \, m\phi &\equiv \begin{cases} 
      1 & m=0 \\
      \sqrt{2} \cos \, m\phi & m > 0 \\
      \sqrt{2} \sin \, |m|\phi & m < 0 
   \end{cases}
\end{split}
\end{align}

where $\Tilde{P}^m_l(cos\theta)$ are normalised associated Legendre polynomials, $l_{\rm max}$ is the meridional order of expansion, $m_{\rm max} \leq l_{\rm max}$ is the azimuthal order of expansion and $m_0(l) = \max(l, m_{\rm max})$. The zeroth order term $(l=0,m=0)$ is a spherical potential. All of the multipole potentials are produced using the \texttt{multipole} function in \textsc{Agama}. In this work, we tested expansions up to a variety of azimuthal orders, but in this report we focus on the $m_{\rm max} = 0$ (axisymmetric) and the $m_{\rm max} = 2$ (quadrupole) cases. In both potentials we allow up to $l_{\rm max} = 6$. The difference between orbits in an $m_{\rm max} = 2$ and in an $m_{\rm max} > 2$ potential was found to be negligible, and so only the axisymmetric and quadrupole potentials are considered in the rest of this report. To quantify the non-axisymmetric of the potential, we calculate the power spectrum $C_l$, shown in Table~\ref{tab:coefficients}, which are calculated via

\begin{equation}\label{eq:power_coeff}
   C_l(r) = \frac{1}{2l+1}\sum_{m=-m_0(l)}^{m=m_0(l)}|\Phi_{l,m}(r)|^2, 
\end{equation}

where $\Phi_{l,m}(r)$ are the multipole coefficients calculated at radius $r$.

\subsection{The Large Magellanic Cloud}

In additional to the static merger remnant potential, we add additional components to simulate the infall of the LMC, and its effect on the merged satellite debris. To simulate the LMC infall, we follow the method described in \citet[][]{vasiliev2021tango}, whereby the reflex motion of the host toward the LMC is accounted for. Since the merger remnant mass is slightly less than the actual present value of the MW, we reduce the mass of the LMC by 4/5 relative to the value found by \citet[][]{erkal2019total}, and adjust the respective scale radii and outer-cutoff radii appropriately. Our LMC model contributes two additional potential components: a direct gravitational potential and a uniform time-dependent acceleration resulting from the reflex of the merger remnant towards the LMC. The equations of motion for the trajectory of the LMC and MW \citep[Eq.~13 in ][]{davies2023ironing} are integrated backwards in time, from the present, for 3 Gyr. The result is used to used to compute the uniform acceleration potential that the satellite stellar debris is integrated in, for 3 Gyr up to the present. The visual illustration of the $(x,y)$ trajectory of the LMC, relative to the merger debris, can be seen in Fig.~\ref{fig:debris}.

\section{Striations in merger debris}\label{sec:mergerdebris}

In this section, we report findings about the effect of our LMC model on the merger debris in an \textit{axisymmetric} multipole potential. We identify an imprint of the LMC in the angular momentum distribution of the debris, which manifests as the appearance of vertical striations in $(L_z, E)$ space. We analyse these features and explain their formation. 

\subsection{Vertical striations}

\begin{figure}
    \centering
    \includegraphics[width=0.75\columnwidth]{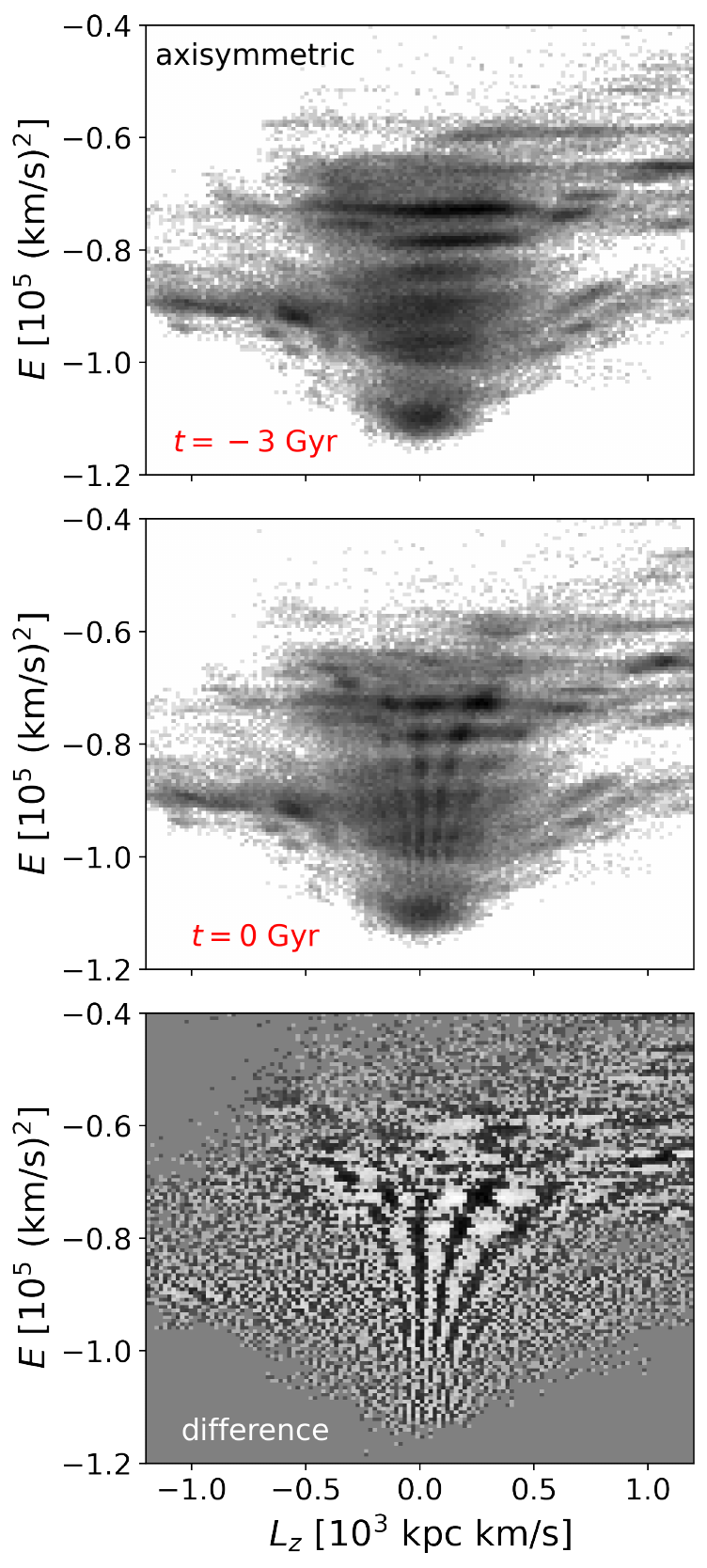}
    \caption{Comparison of the $(L_z, E)$ distribution of the satellite merger debris before and after the LMC interaction, in a axisymmetric host potential. The top panel shows the debris after 5 Gyr of phase mixing, whereas the bottom panel shows the debris after a further 3 Gyr of evolution, with the LMC. Note the appearance of the new vertical striations present at all energy values but focused, and roughly symmetric, about $L_z = 0$. The appearance of these lines is most visible when the LMC is at its pericentre of about 50 kpc.}
    \label{fig:compare_lines}
\end{figure}

After integrating the merger debris for 3 Gyr in the combined (axisymmetric) host and LMC potential, we compare the values of energy $E$ and z-angular momentum $L_z$ of the final snapshot (at the present day) to those 3 Gyr in the past. In Fig.~\ref{fig:compare_lines} we present three panels that show 2-d log-density $(L_z, E)$ distributions at the aforementioned times, as well as the difference between the two, to make the features more clear. At the LMC's closest approach, there are 3 main vertical features which persist all the way from the minimum energy value up to about $E \simeq 0.5\times 10^{5}$ (km/s)$^2$. There may be additional features, but the low particle density outside of $|L_z| > 0.5$ kpc km/s makes them difficult to see. Note also that the vertical striations are compact at low energy and subsequently spread further apart at higher energies. Given that the merger debris is already heavily substructured in energy, the LMC induced vertical substructures create new pockets of substructure throughout the $(L_z, E)$ distribution. 

\begin{figure}
    \centering
    \includegraphics[width=0.8\columnwidth]{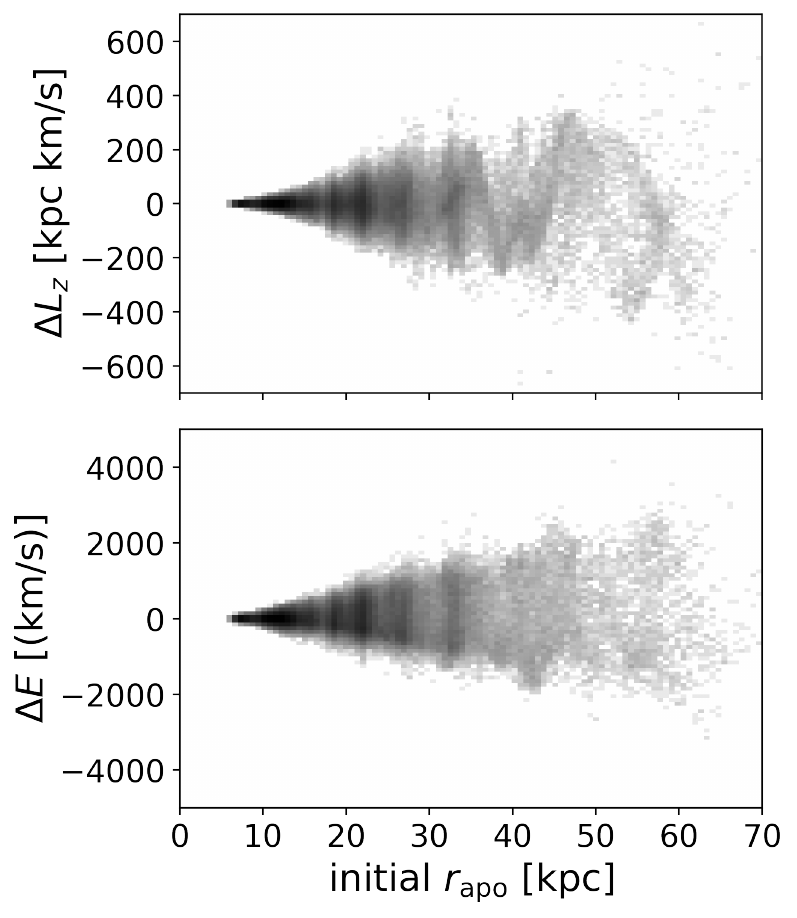}
    \caption{The change in energy and angular momentum vs initial apocentre for all particles in the simulation. We see that the change in angular momentum is far more substantial and the change in energy. Additionally, a series of zig-zag features are present in the $\Delta L_z$ plot, but not in energy. Motivated by the relationship of $r_{\rm apo}$ to other orbital parameters, we can look for a periodic link between $\Delta L_z$ and some other orbital property, like orbit orientation or frequency. }
    \label{fig:r_apo}
\end{figure}

To understand the cause of these vertical striations, we investigate the change of energy and angular momentum versus some key orbital parameters. In Fig.~\ref{fig:r_apo}, we show $\Delta L_z$ and $\Delta E$ against the initial value of apocentric radius $r_{\rm apo}$ for all particles in the merger debris. Here $\Delta E$ refers to the value of $E$ at the present day compared with 3 Gyr before the present, when the LMC was sufficiently far away from the Galactic centre to have no impact (similarly for $\Delta L_z$). Most evident is the fact that both $\Delta E$ and $\Delta L_z$ increase as initial $r_{\rm apo}$ increases. We find that, for this model of the LMC, $r_{\rm apo}$ determines the amount by which integrals of motion will change more so than any other distance measurement like distance from the LMC, for example. That is to say, the outer stellar halo is the most affected by the LMC. Fig.~\ref{fig:r_apo} also makes clear that a change in $L_z$ is more likely the cause of vertical striations than a change in $E$. We see this because both $\Delta L_z \sim \mathcal{O}(L_z)$ kpc km/s and $\Delta E \sim \mathcal{O}(10^{-2} E)$ (km/s)$^2$. Moreover, we notice an unusual zig-zag pattern forming at $r_{\rm apo} \gtrsim 30$ kpc for $\Delta L_z$, while the energy distribution contains no such feature. 

\begin{figure}
    \centering
    \includegraphics[width=0.8\columnwidth]{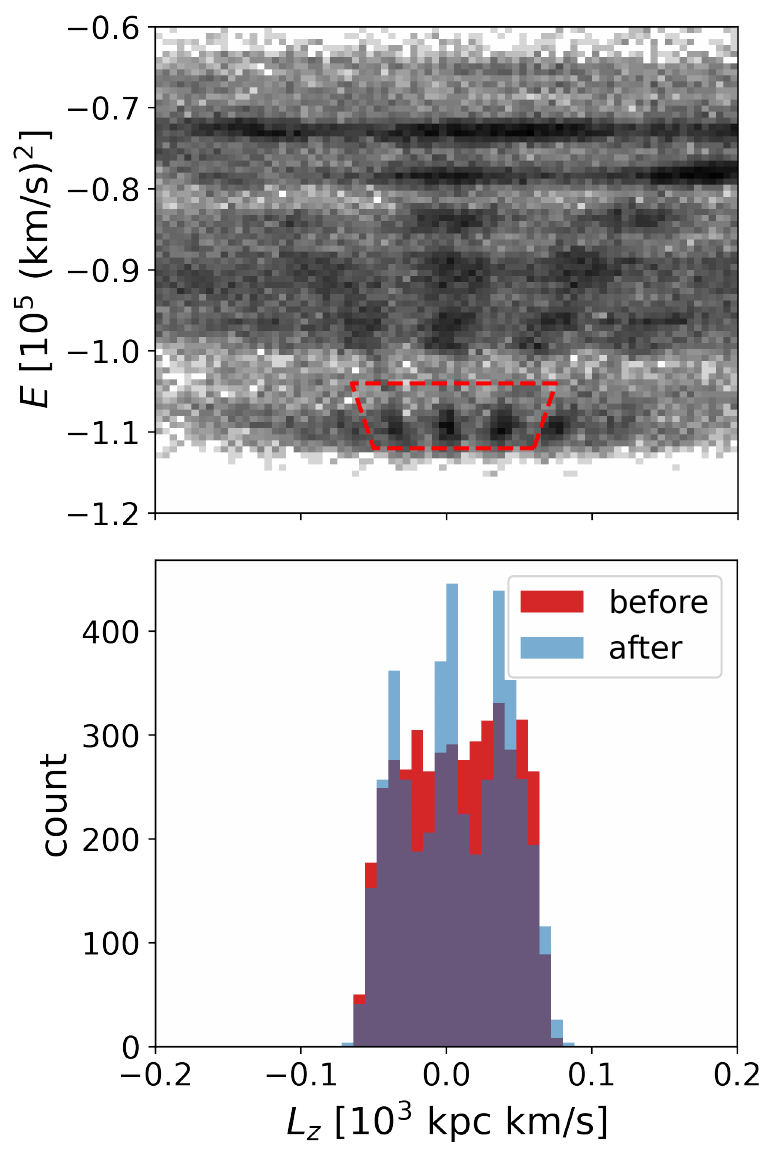}
    \caption{A zoomed-in view of the $(L_z, E)$ distribution, where we have selected the low-energy portion of the three vertical striations in a red dashed trapezoid. The bottom panel reveals how these three lines were not present in the original distribution by way of a 1-d histogram of the z-angular momenta selected above. The red distribution shows the values of $L_z$ present 3 Gyr before the LMC's pericentric passage, while the blue distribution shows these values at the point where the LMC is at pericentre. Three vertical peaks have clearly formed.}
    \label{fig:lines_zoom_in}
\end{figure}

We now isolate a number of particles which belong to a low energy region of the vertical striations. In the top panel of Fig.~\ref{fig:lines_zoom_in}, we illustrate the selected particles by a dashed red trapezoid in $(L_z, E)$ space. Note that we have zoomed-in relative to Fig.~\ref{fig:compare_lines}, and show a much more narrow range of $L_z$ values. In the bottom panel, we plot a 1-d histogram of the selected particles $L_z$ values at 3 Gyr before the LMC closest approach (``before'') and at the present day (``after''). It is clear to see that an initially quite uniform distribution of $L_z$ values is transformed into three distinct peaks once the LMC approaches its pericentre. These peaks (corresponding to the vertical striations) have come from particles within the initial range of $L_z$ and fall along three very distinct ridges.

Given that these features are present only in $L_z$, and no other component of angular momentum, it is worth noting the orientation of the merger debris relative to the trajectory of the LMC. In Fig.~\ref{fig:debris}, we present $(x,y)$ projection of both the entire merger debris (in black) as well as the selected debris (in red). The debris is set so that the initial orientation of the long axis is set at 70 degrees from the $x$-axis, in accordance with \citet[][]{iorio2019shape}. The trajectory of the LMC is seen as a dashed red line, with its final position (at the present day) shown by an open red circle.

\subsection{Precession of near radial orbits}

\begin{figure}
    \centering
    \includegraphics[width=0.8\columnwidth]{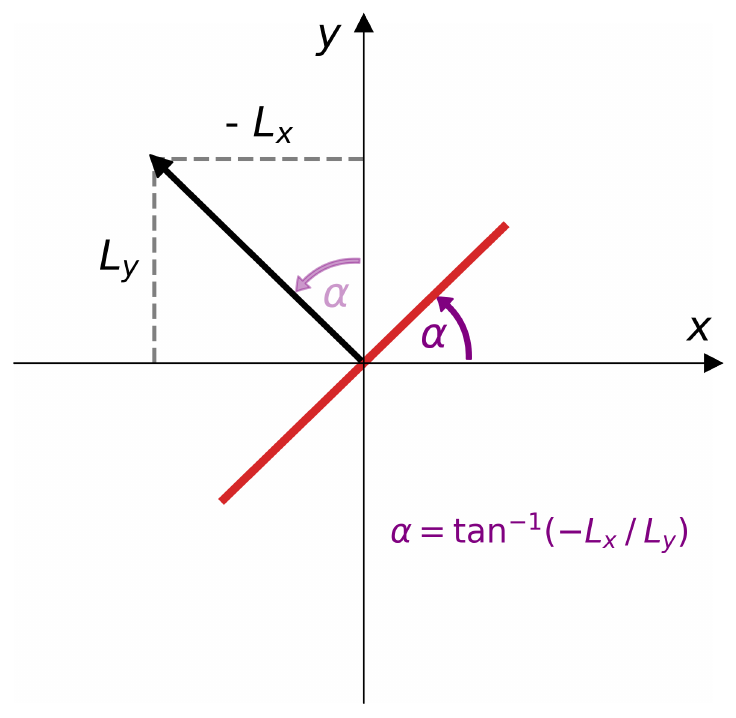}
    \caption{Diagram illustrating the physical significance of the orbital orientation angle $\alpha$, which is defined by the equation in purple text. The thick red line shows the trajectory of an approximately polar orbit, and the perpendicular black arrow is the $(x,y)$ projection of this orbit's angular momentum vector $\bm{L}$. The angle $\alpha$ measures the orientation of such an orbit by taking the inverse tangent of the ratio of $-L_x$ and $L_y$.}
    \label{fig:alpha_diagram}
\end{figure}

\begin{figure}
    \centering
    \includegraphics[width=0.8\columnwidth]{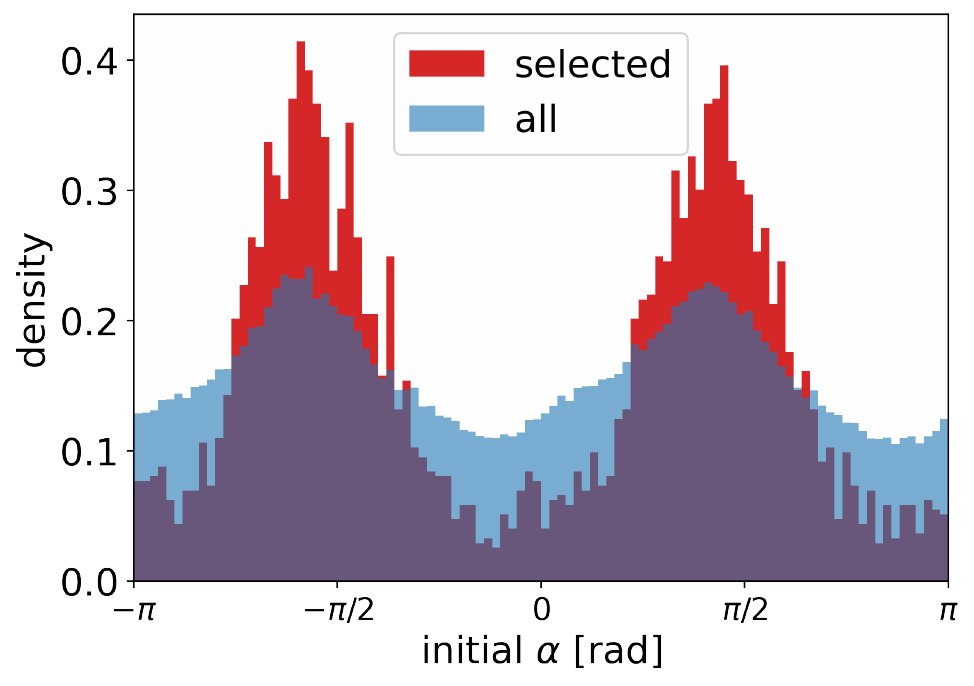}
    \caption{The initial ($t=-3$ Gyr)  $\alpha$ distribution for both the selected particles, and for the entire population of satellite debris. The extremely narrow distribution for the selected region makes the effect very noticeable. However, there is still a sufficiently non-uniform distribtion of $\alpha$ in the entire population such that the striation effect is noticed at all energy values.}
    \label{fig:merger_alpha_dist}
\end{figure}

With the appearance of these features made clear, we now turn to a phenomena that contributes to explaining the formation of these features -- orbital plane precession of near radial orbits. In brief, the initial (when the LMC is far away) value of $L_z$ determines the rate of orbital precession about the $z$-axis for near radial orbits. Therefore, a periodic relationship is established between the initial $L_z$ and the orientation of near radial orbits at the time of the LMC's infall. We introduce a parameter $\alpha$ that defines the physical orientation of a particle on an approximately radial orbit in the $(x,y)$ plane:
\begin{equation}
    \alpha = \arctan\left(\frac{-L_x}{L_y}\right),
\end{equation}
where $L_x$ and $L_y$ are the projections of the angular momentum vector onto the $x$ and $y$ axis respectively. Radial orbits can be visualized as having two apocentres on opposite sides of the galaxy, orientated by some angle \citep[Section 5.5 of][]{binneyandtremaine2008}. The angle $\alpha$ therefore gives the orientation of a nearly radial orbit with respect to the $x$-axis, as shown diagrammatically in Fig.~\ref{fig:alpha_diagram}. In this figure, the red bar represents such an orbit that is orientated at an angle $\alpha$ form them the $x$-axis. We use this orbital orientation angle to quantify the physical spread of near radial orbits in $(x,y)$ plane. In Fig.~\ref{fig:merger_alpha_dist}, we present the 1-d histogram of initial ($t=-3$ Gyr) $\alpha$ values for all the particles in the simulation (in blue) alongside the initial $\alpha$ values for the selected particles. In both cases, we see a non-uniform distribution, with an especially narrow distribution for the selected particles between $\alpha \simeq \pi/4$ and $\alpha \simeq \pi/2$. These narrow ranges of $\alpha$ values visually agree with physical orientation that is seen in Fig.~\ref{fig:debris}, where we can see that both populations of debris particles are quite narrowly distributed along a line at 70 degrees to the $x$-axis. This initially narrow range of $\alpha$ values, resulting from the nature of the GSE merger, will turn out to be key to the appearance of the vertical striations.

In Fig.~\ref{fig:alpha_time}, we present how the value of $\alpha$ at a given time step depends on the initial angular momentum of a particle in a simulation without the presence of LMC (i.e. in a static axisymmetric potential). We plot only the selected particles. From left to right, we show the value of $\alpha$ at four different time snapshots. The initially narrow band of $\alpha$ begins to wind up as orbits precess at varying rates. Since the value of $L_z$ of an orbit determines its rotation about the $z$-axis, $L_z$ determines the precession of an orbit in the $(x,y)$ plane. An orbit with a positive value of $L_z$ will precess anticlockwise in the $(x,y)$ plane, while an orbit with a negative value of $L_z$ will precess clockwise in the $(x,y)$ plane. After some time, the initially narrow band of $\alpha$ results in a periodic dependence between $\alpha(t)$ and $L_z$, as shown in the right-most panel of Fig.~\ref{fig:alpha_time}. 

\subsection{Striation Mechanism}

\begin{figure}
    \centering
    \includegraphics[width=0.75\columnwidth]{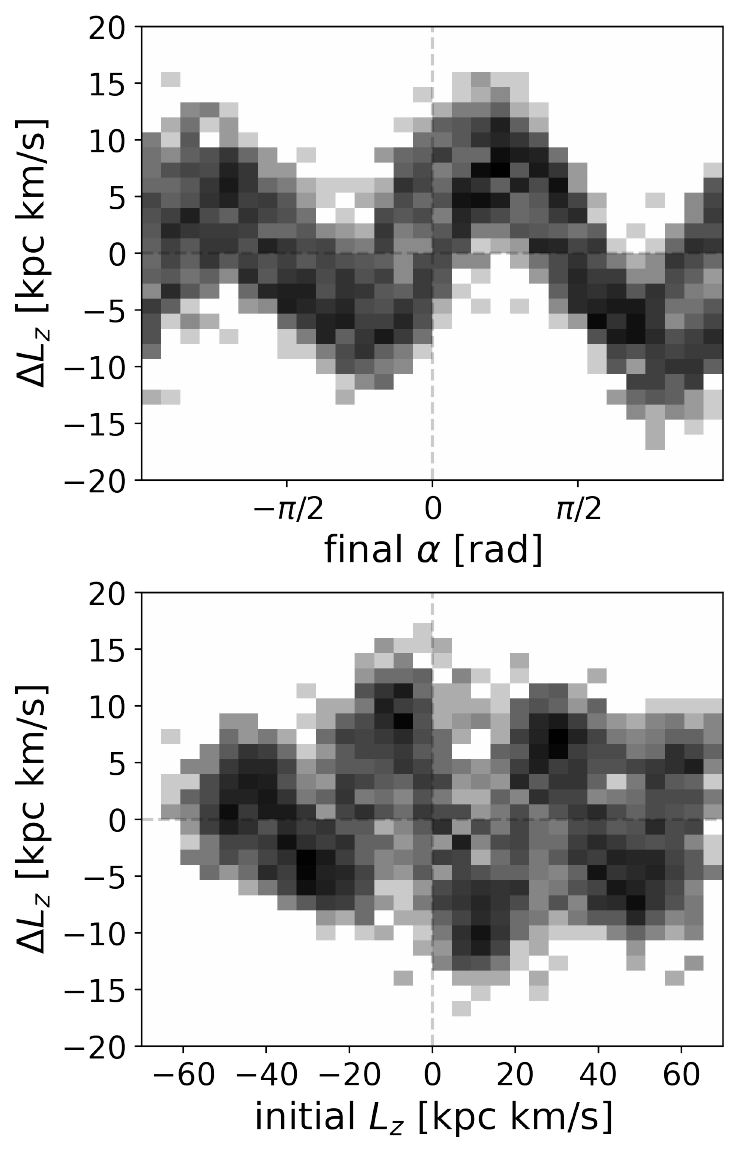}
    \caption{The periodic dependence of $\Delta L_z$ on $\alpha$, and therefore on initial $L_z$, for the selected particles in the merger debris. The periodic dependence on $\alpha$ is diagrammatically explained in Fig.~\ref{fig:alpha_diagram}. Note the peaks at $\alpha \sim \pi/4$ and $-3\pi/4$, and the troughts at $\alpha \sim 3\pi/4$ and $-\pi/4$. We attribute these peaks and troughs to the orientation of the orbits relative to the LMC's trajectory.  Since the initial value of $L_z$ dictates the orientation $\alpha$, and the value of $\alpha$ determines the value of $\Delta L_z$, we find a periodic relationship between $L_z$ and $\Delta L_z$, illustrated in the bottom panel. This periodicity causes the formations of the lines in $L_z$, as seen originally in Fig.~\ref{fig:compare_lines}.}
    \label{fig:periodic}
\end{figure}

Since a periodic relationship has been established between $\alpha$ and initial $L_z$, it makes sense to look for a relationship between $\alpha$ and $\Delta L_z$, given that we are sure that the change in $L_z$ is driving the formation of the vertical striations. In the top panel of Fig.~\ref{fig:periodic} we plot $(\alpha,\Delta L_z$) and see a clear periodic relationship between the final value of $\alpha$ and $\Delta L_z$ for all of the selected particles. This periodicity between $\alpha$ and $\Delta L_z$, combined with the relationship between $\alpha$ and initial $L_z$ implies a relationship between initial $L_z$ and $\Delta L_z$. This is confirmed in the bottom panel of Fig.~\ref{fig:periodic}, where we see a repeating pattern of positive and negative $L_z$ as we move across from large negative initial $L_z$ to large positive initial $L_z$. 

\begin{figure*}
    \centering
    \includegraphics[width=0.95\textwidth]{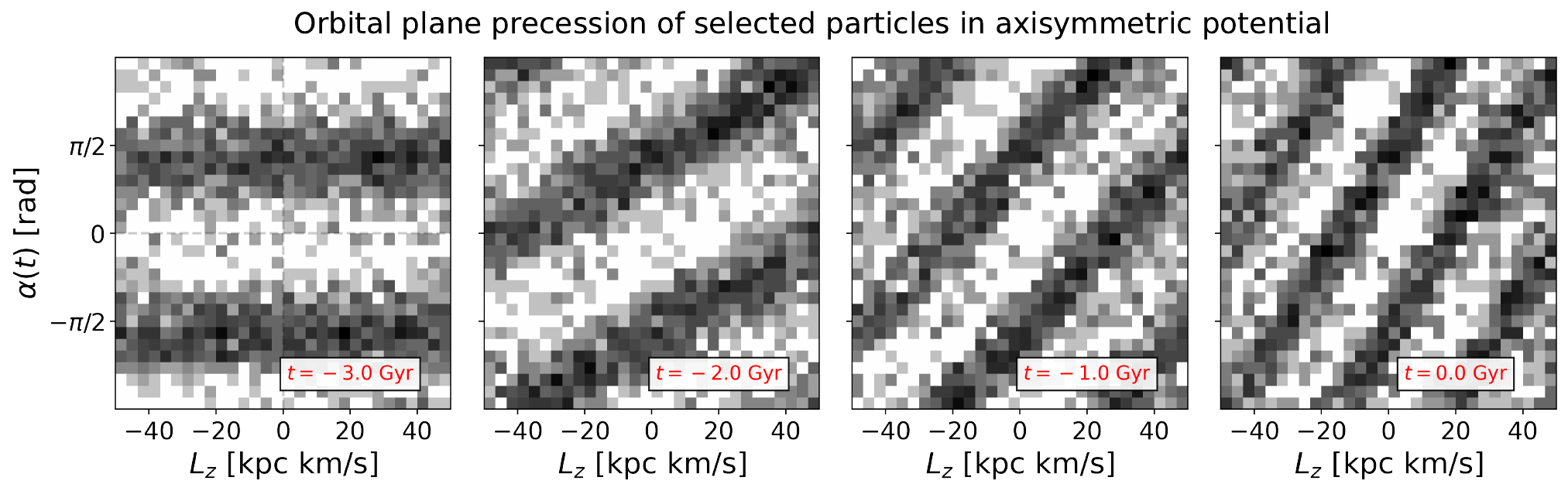}
    \caption{The distribution of orbital orientation angle $\alpha$ against z-angular momentum over time, in the absence of the LMC, for the selected particles in the merger debris. Note the initially narrow distribution of $\alpha$, which begins to wind up over time. Since the value of $L_z$ dictates the precession rate of these orbits, eventually we are left with a periodic dependence between $\alpha$ and $L_z$.}
    \label{fig:alpha_time}
\end{figure*}

\begin{figure}
    \centering
    \includegraphics[width=0.8\columnwidth]{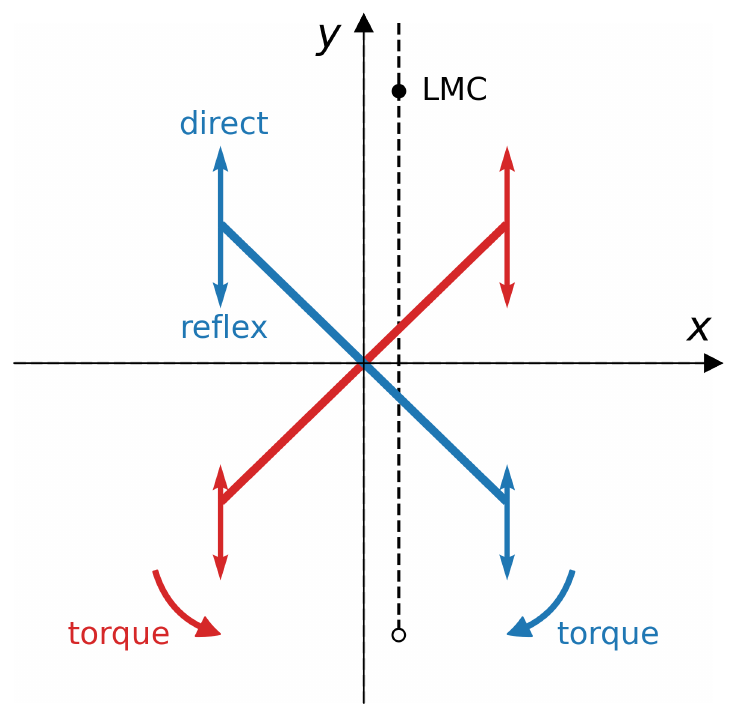}
    \caption{Schematic illustrating how the imbalance of forces acting on approximately polar orbits creates a torque, either increasing or decreasing the value of $L_z$. Since the reflex force is the same for all particles, the torque is generated primarily because of the difference between the direct forces. The direct force from the LMC is stronger for particles in close proximity to it, which causes the orbit's $L_z$ vector to change. From this diagram, one expects positive torque at $\alpha \sim \pi/4$ and $-3\pi/4$, and negative torque at $\alpha \sim 3\pi/4$ and $-\pi/4$.}
    \label{fig:force_diagram}
\end{figure}

The relationship between $\Delta L_z$ and $\alpha$ does not have such a simple explanation as the $(L_z,\alpha)$ pattern, and therefore requires some conjecture. We provide Fig.~\ref{fig:force_diagram} as a visual aid for our explanation. Additionally, in the following subsection, we provide a toy mathematical model to justifying our explanation.

Firstly, note that from the top panel of Fig.~\ref{fig:alpha_time}, we see that particles have their angular momentum changed the most at $\alpha \sim \{\pi/4, 3\pi/4, -\pi/4, -3\pi/4\}$. Since the LMC's trajectory is approximately along the $y$-axis (see Fig.~\ref{fig:debris}), as it approaches from the large positive $y$ the direct force toward the LMC will be stronger for particles with $\alpha = 0^+$ to $\pi/2$ than for particles with $\alpha = 0^-$ to $-\pi/2$. While as the LMC moves past the origin and toward large negative $y$ values the direct force will be stronger for particles with $\alpha = 0^-$ to $-\pi/2$ than for particles with $\alpha = 0^+$ to $\pi/2$. 

Moreover, the near radial orbits have orbital periods sufficiently shorter than the LMC infall time. For the selected particles shown in Fig.~\ref{fig:lines_zoom_in}, the median radial period is sufficiently small at 0.21 Gyr. The median of the entire population of the merger debris is 0.45 Gyr. These fast radial frequencies, combined with the imbalance of forces will be create a torque in the $z$-direction ($\tau_z$), that will be maximum for particles with $\alpha \sim \{\pi/4,3\pi/4\}$ when the LMC approaches from positive $y$ values, and will be maximum for particles with $\alpha \sim \{-\pi/4,-3\pi/4\}$ as the LMC approaches toward negative $y$ values. For particles with $\alpha \sim \{-\pi/2, 0, \pi/2, \pi\}$, this torque will be minimal due to the orientation of the LMC's trajectory, as the particles complete fast near radial orbits such that forces balance out on either side. In a sense we can consider the near radial orbits as as a rigid wire experiencing a force from the LMC. The subsequent LMC induced torque will therefore cause a change to $L_z$, which has maximum effect at $\alpha \sim \{-3\pi/4, -\pi/4, \pi/4, 3\pi/4\}$. We see this quite clearly in the top panel of Fig.~\ref{fig:periodic}. 

\begin{figure*}
    \centering
    \includegraphics[width=0.9\textwidth]{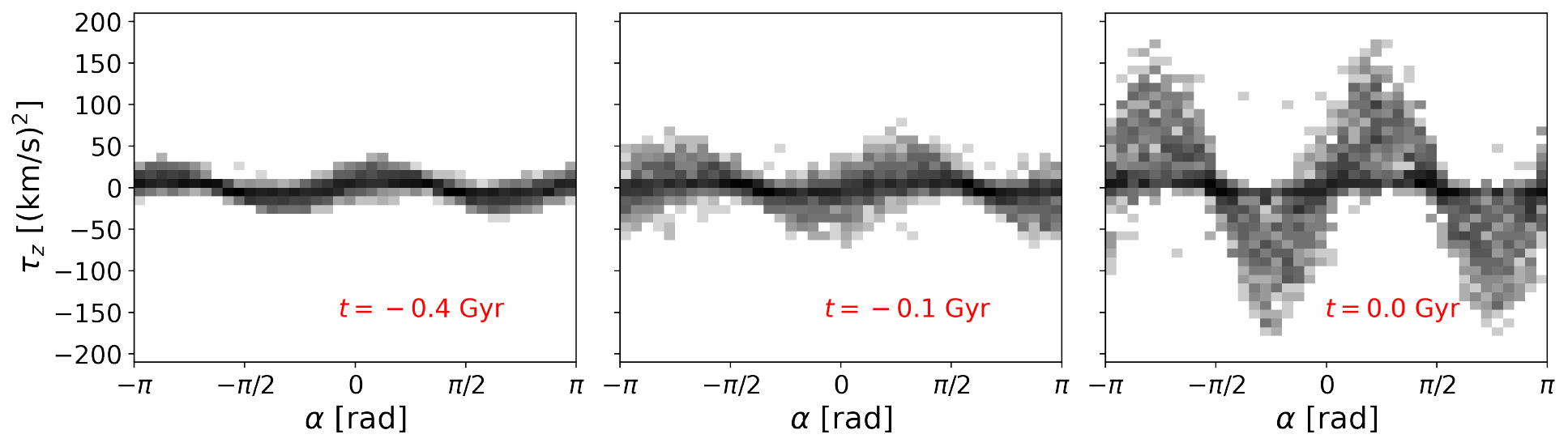}
    \caption{The value of torque in the $z$-component, $\tau_z$, against orbital orientation angle, $\alpha$, for three time snapshots. Note we present the values only for the selected particles. We see positive total $\tau_z$ for $\phi \sim \pi/4$ and $-3\pi/4$, and negative total $\tau_z$ for $\phi \sim 3\pi/4$ and $-\pi/4$. This corresponds to the change in $L_z$ for the corresponding values of $\alpha$. Compare the values of $\alpha$ for which the torque is maximum and note that it matches with the $\alpha$ for which the $\Delta L_z$ is maximum in Fig.~\ref{fig:periodic}, confirming our hypothesis that was presented diagrammatically in Fig.~\ref{fig:force_diagram}.}
    \label{fig:torque}
\end{figure*}

We provide an additional illustration of the torque, $\tau_z$, of all particles in the simulation in Fig.~\ref{fig:torque}. In this figure, we show how the magnitude of $\tau_z$ is greatest at $\alpha \sim \{-3\pi/4, -\pi/4, \pi/4, 3\pi/4\}$, which corresponds to the values of $\alpha$ in Fig.~\ref{fig:periodic}. Additionally, the amplitude of the torque increases as we approach the present time i.e. as the LMC gets closer to its pericentre.

\subsection{Toy model of LMC torquing}

In this section, we provide a toy mathematical description of the way in which a large body, such as the LMC, causes the angular momenta of nearly radial orbits to change. Consider a group of particles all on similar (near radial) orbits, with the same value of $\alpha$. If we approximate these particles as a rigid homogeneous bar with mass $m$ and length $2a$, and the LMC as a point mass $M$ some distance $r$ away, we can calculate the force (and the torque) of this toy LMC on the rigid bar.  We can therefore consider $a$ as the apocentric distance of these orbits. In the presence of a perturber, this approximation holds best in the adiabatic limit, where the radial periods of the particles are much smaller than that of the perturber. 

\begin{figure}
    \centering
    \includegraphics[width=0.75\columnwidth]{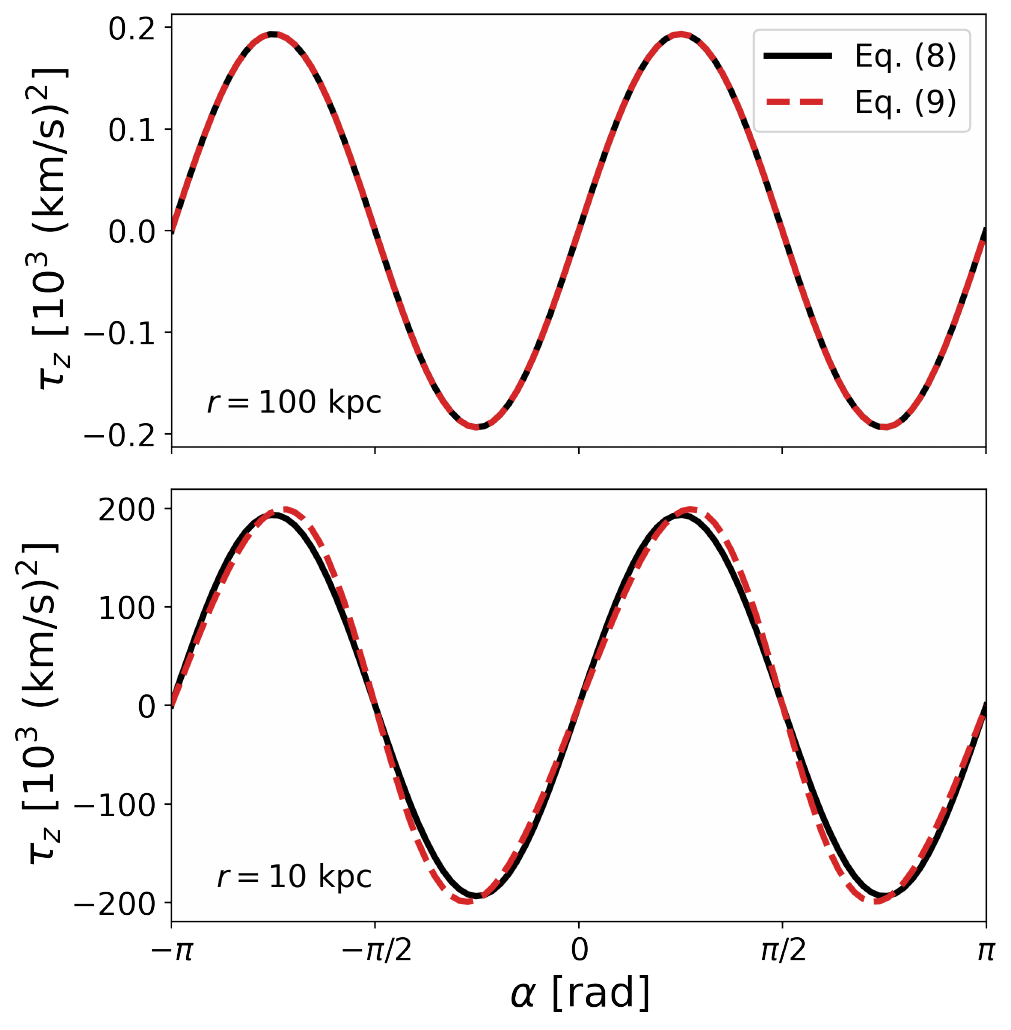}
    \caption{The torque resulting from the gravitational force between a point mass (representing the LMC) and a rigid homogeneous bar (representing a near radial orbit). We plot torque as a function of $\alpha$, using both Eq.~\ref{eq:full_torque} (in black) and Eq.~\ref{eq:lim_torque} (in red), where $M = 10^{11}$ M$_{\odot}$, $m = 10^{2}$ M$_{\odot}$ and $a = 3$ kpc. In the top panel we take the distance of the point mass from the centre of the bar to be $r=100$ kpc, and in the bottom panel we take $r=10$ kpc. Note the maxima at $\alpha = \{-3\pi/4, \pi/4\}$ and minima at $\alpha = \{-\pi/4, 3\pi/4\}$, which matches with the shape of the orbital orientation angles that experience the maximum torque in Fig.~\ref{fig:torque}.}
    \label{fig:model_torque}
\end{figure}

In this model we assume that the motion of the rigid bar and the point mass is restricted to the $(x,y)$ plane. Moreover, assume that the centre of the rigid bar is at $(x=0,y=0)$ and that the bar is at an angle $\alpha$ from the $x$-axis. The point mass lies some distance $r$ away from the centre of the bar at $(x=0, y=r)$. At this distance, the point mass exerts a force $d\bm{F}$ on each mass element $dm$ of the bar given by 
\begin{equation}
    d\bm{F} = \frac{GMdm}{d^2}\bm{n},
\end{equation}
where $\bm{n}$ is the vector pointing toward the point mass, and $d^2$ is the distance of point mass to $dm$. If we let $\lambda$ be the coordinate along the length of the bar (i.e. in the $\bm{\hat{n}_1}$ direction), and let $dm = (m/2a)d\lambda$, then this becomes
\begin{equation}
    d\bm{F} = \frac{GMmd\lambda}{2a} \frac{ (r\sin(\alpha) - \lambda)\bm{\hat{n}_1} + r\cos(\alpha)\bm{\hat{n}_2}}{\left[ r^2 \cos^2(\alpha) + (r\sin(\alpha) - \lambda)^2\right]^{3/2}} 
\end{equation}
where we have defined new coordinates $(\bm{\hat{n}_1}, \bm{\hat{n}_2})$ with $\bm{\hat{n}_1}$ along the bar's length and $\bm{\hat{n}_2}$ perpendicular to the bar. The relationship between the two coordinate systems is given by
\begin{equation}
    \begin{pmatrix}
    \bm{\hat{n}_1} \\
    \bm{\hat{n}_2}
    \end{pmatrix}
    =
    \begin{pmatrix}
    \cos\alpha & \sin\alpha \\
    -\sin\alpha & \cos\alpha
    \end{pmatrix}
    \begin{pmatrix}
    \bm{\hat{x}} \\
    \bm{\hat{y}}
    \end{pmatrix},
\end{equation}
whereas the $\hat{\bm{z}}$ axis is unchanged. To calculate the torque, we then integrate $\bm{r} \times d\bm{F}$ over the length of the bar:
\begin{equation}
    \bm{\tau} = \int_{-a}^a \bm{r} \times d\bm{F}
\end{equation}
where $\bm{r} = \lambda\bm{\hat{n}_1}$ is along the bar's axis. The torque integral then simplifies to
\begin{equation}\label{eq:full_torque}
    \bm{\tau} = \hat{\bm{z}}\frac{G M m }{2a} \int_{-a}^a \frac{r\cos(\alpha)\lambda d\lambda}{\left[ r^2 \cos^2(\alpha) + (r\sin(\alpha) - \lambda)^2\right]^{3/2}}.
\end{equation}
This results in a reasonably unwieldy expression that, in the limit of $r \gg a$, simplifies to
\begin{equation}\label{eq:lim_torque}
    \bm{\tau} = \frac{GMma^2}{2r^3}\sin(2\alpha)\hat{\bm{z}},
\end{equation}
which has maxima when $\alpha = \{-3\pi/4, \pi/4\}$ and minima when $\alpha = \{-\pi/4, 3\pi/4\}$. In Fig.~\ref{fig:model_torque}, we plot both Eq.~\ref{eq:full_torque} and Eq.~\ref{eq:lim_torque} as a function of $\alpha$, and see this confirmed. We solve Eq.~\ref{fig:model_torque} using \texttt{scipy.integrate.quad} and take values of $M = 10^{11}$ M$_{\odot}$, $m = 10^{2}$ M$_{\odot}$ and $a = 3$ kpc. Note also that $\tau \sim r^{-3}$, showing that the torque increases as the particle approaches the bar. Both equation Eq.~\ref{eq:full_torque} and Eq.~\ref{eq:lim_torque} diverge at $r=0$, but a simple extension to 3-d fixes this if the point mass is held at some fixed non-zero $z$.

\section{Comparison between potentials}\label{sec:compare}

In this section we compare the strength of the vertical striations in two types of static multipole potentials, which are generated from all particles in the $N$-body snapshot after 5 Gyr. In the previous section we had considered only an axisymmetric potential, where we now compare this with a bisymmetric quadrupole potential. We note that the difference between the axisymmetric and quadrupole potentials is small, yet the change in orbits is significant. As a sort of sanity check, we compare some aspects of the test particle simulations with a version of the $N$-body simulation which was ran for a full 8 Gyr.

\subsection{Shape of the potentials}

\begin{figure*}
    \centering
    \includegraphics[width=0.93\textwidth]{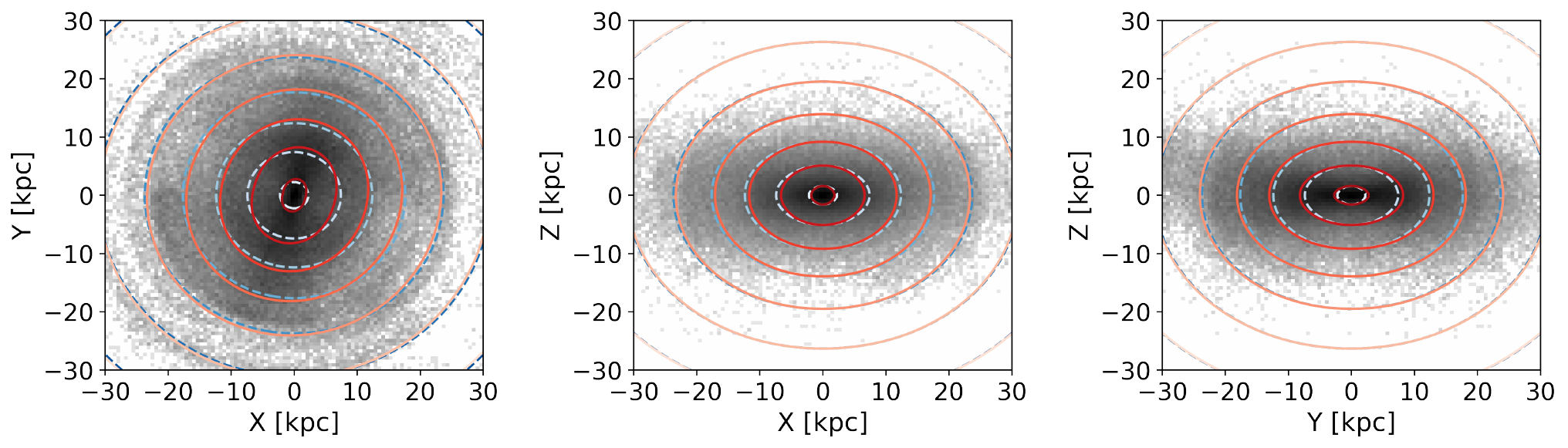}
    \caption{Comparison between the axisymmetric potential and bisymmetric quadrupole potential of the satellite merger debris (DM + stellar particles). The red contour lines show the shape of the quadrupole potential, the grey dashed contour lines show the shape of axisymmetric potential and the log density plot of the stellar satellite debris (at $t=-3$ Gyr) is shown in black. This figure makes it more clear that the quadrupole potential is only weakly non-axisymmetric.}
    \label{fig:potential_compare}
\end{figure*}

Because the mass of the merger remnant is dominated by the host, the axisymmetric and quadrupole potentials formed from the entire merger debris (DM + stellar particles) only differ very slightly on large scales; the difference is only noticeable in the inner regions of the potential. In Fig.~\ref{fig:potential_compare} we show an axisymmetic and bisymmetric quadrupole potential, generated from \textit{only} the satellite merger debris (DM + stellar particles), to emphasise the difference between the two. In Fig.~\ref{fig:potential_compare}, the merged satellite debris is shown in black as a 2-d log density histogram, the equipotential contour lines of the quadrupole potential are shown in red, and the contours of the axisymmetric potential are shown as dashed grey lines. The difference is most clear in the inner $10$ kpc, where we an see that the potential is squashed along the direction of the debris major axis. Additionally, the potential looks slightly squashed in the $z$ direction. The major axis of the debris, and therefore the quadrupole potential is at 70 degrees from the $x$ axis by construction, as described previously.

To quantify the difference between the quadrupole potential and the axisymmetric potential, we take a similar approach to the way in which \citet[][]{dehnen2000effect} quantifies the strength of the bar (quoted as the dimensionless parameter $\alpha$). That is, for the total potential, we calculate the ratio of the force due to the quadrupole component and the axisymmetric background at some radius, along the major axis of the quadrupole potential. To do this, we calculate the force (along the major axis) in the total quadrupole potential, $f_{\rm quad}$, the force in the total axisymmetric potential, $f_{\rm axi}$, and calculate the ratio of $f_{\rm quad} - f_{\rm axi}$ to $f_{\rm axi}$. We find that at radii of $8$ kpc (the solar radius) along the major axis, the ratio of the forces is $6\times10^{-4}$. Therefore, the contribution of the quadrupole is small when compared with this force ratio found for the bar. For example, in \citet[][]{dehnen2000effect} the ratio of the bar's quadrupole force to the axisymmetric background is taken to be around $10^{-2}$ at 8 kpc. Evidently, these near radial orbits are very sensitive to even very small differences in the potential, and care must be taken in choosing the right potential.

\subsection{Appearance of striations}

\begin{figure}
    \centering
    \includegraphics[width=0.75\columnwidth]{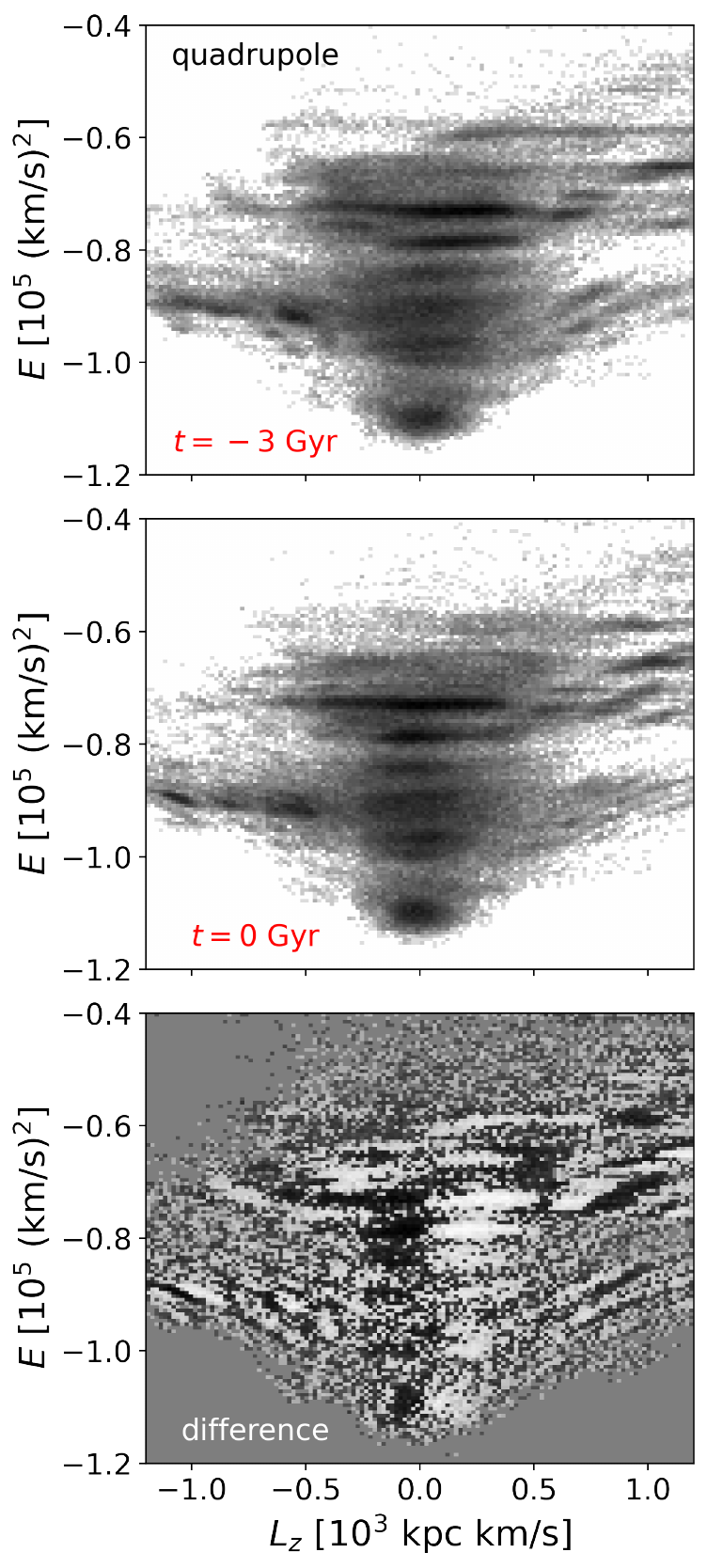}
    \caption{The same as in Fig.~\ref{fig:compare_lines}, but for the quadrupole potential. Note the absence of the vertical striations, except for appearance of some non-vertical striations at greater positive and negative values of angular momentum.}
    \label{fig:compare_lines_quad}
\end{figure}

As in the case of the axisymmetric potential, we examine how the $(L_z,E)$ distribution of the merged satellite debris changes as the LMC approaches its pericentre. In Fig.~\ref{fig:compare_lines_quad} we the show 2-d log-density $(L_z, E)$ distribution 3 Gyr before the LMC is at pericentre and just as the LMC is at pericentre. Additionally, in the bottom panel, we show the difference of these two for emphasis. Note the strong difference between this plot and Fig.~\ref{fig:compare_lines}, where the vertical striations are clearly present at low values of angular momentum. In the quadrupole potential, it appears this striation effect seems to only manifest slightly at large positive and negative values of angular momentum. Additionally, near $L_z \sim 0$, there appears to be an increase in negative values of angular momentum and a decrease in positive values of angular momentum. As discussed below, we assert that the primary reason for the absence of the striations in the quadrupole potential is due to the decreased rate of orbital plane precession.

\subsection{Precession rate in different potentials}

\begin{figure}
    \centering
    \includegraphics[width=0.86\columnwidth]{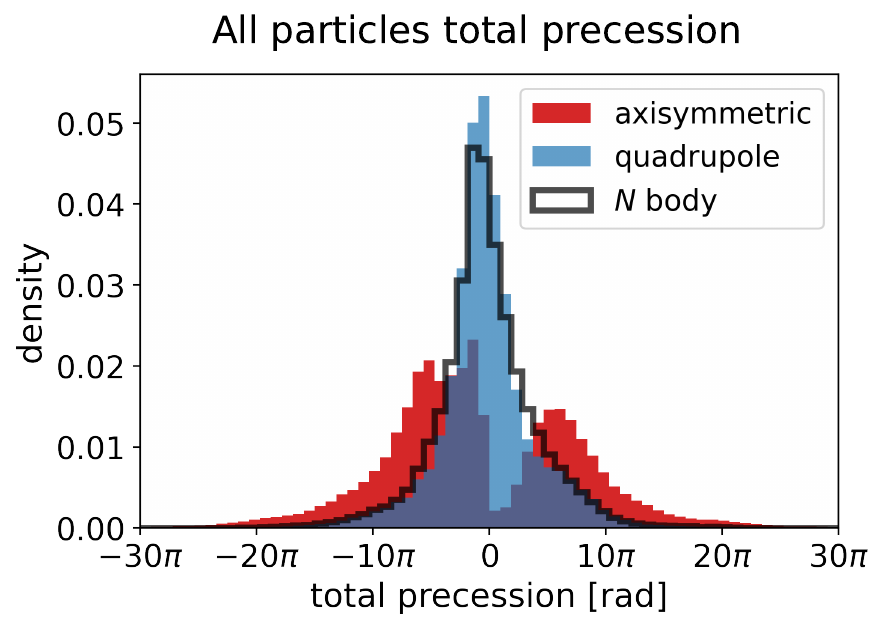} \\
    \hspace{1cm}
    \includegraphics[width=0.98\columnwidth]{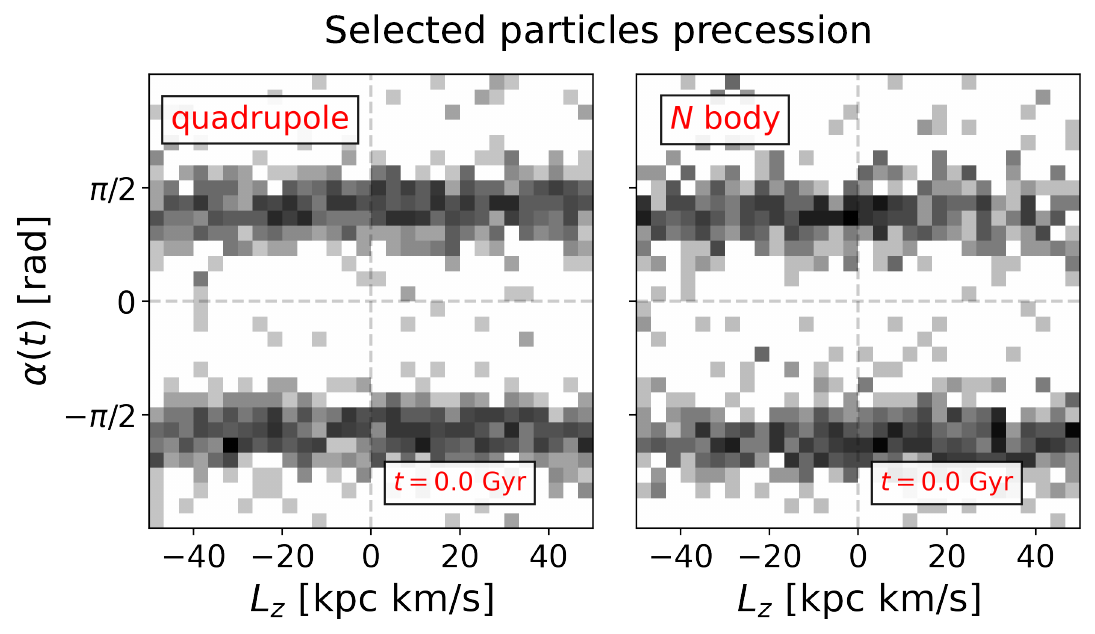}
    \caption{Comparison between the precession rates of the orbital plane angle $\alpha$ for the axisymmetric and quadrupole potentials, and the $N$-body simulation. \textit{Top:} A comparison of the total precession (for all particles in the simulation) in each case. It is clear that the choice of axisymmetric potential causes the test particles to precess at a faster rate than the quadrupole potential. The increased precession rate in the axisymmetric potential causes a stronger periodic relationship between $\alpha$ and initial $L_z$, resulting in the vertical striations. Higher order expansions match up very closely with the quadrupole. \textit{Bottom:} The same as the rightmost panel of Fig.~\ref{fig:alpha_time}, but for the bisymmetric quadrupole potential and the $N$-body simulation. We see here that the quadrupole aligns much more closely with the $N$-body. While there is evidently some precession (from the top figure), it appears as though there is very little for these low energy, low angular momentum selected particles. }
    \label{fig:total_precession}
\end{figure}

In the axisymmetric potential, the cause of the striations was a result of two phenomena: (a) a periodic relationship between orbital precession (measured by $\alpha$) and angular momentum, and (b) a relationship between orbital orientation and change in angular momentum. Since the latter is simply a result of the orientation of the LMC with respect to the debris, it is the same regardless. However, the former relationship is linked to the shape of the potential. To test this, in the top panel of Fig.~\ref{fig:total_precession} we plot a histogram of the total precession (after 3 Gyr) of all the particles in the axisymmetric potential (in red), all the particles in the quadrupole potential (in blue), and all the particles in the 8 Gyr $N$-body simulation (in black outline). We see from this figure that the maximum total precession reached by particles in the axisymmetric potential is much greater than in the quadrupole potential. This suggests that the rate of orbital precession is much greater in the axisymmetric potential. Interestingly, there are very few particles that undergo no precession in the axisymmetric case, but many in the quadrupole case. This substantial change in the orbital precession, despite the only mild inner difference between the merger debris potentials, suggests that near radial orbits are extraordinarily sensitive to the shape of the very inner potential. Note that the quadrupole potential reassuringly matches up particularly well with the 8 Gyr $N$-body simulation.

\begin{figure}
    \centering
    \includegraphics[width=0.65\columnwidth]{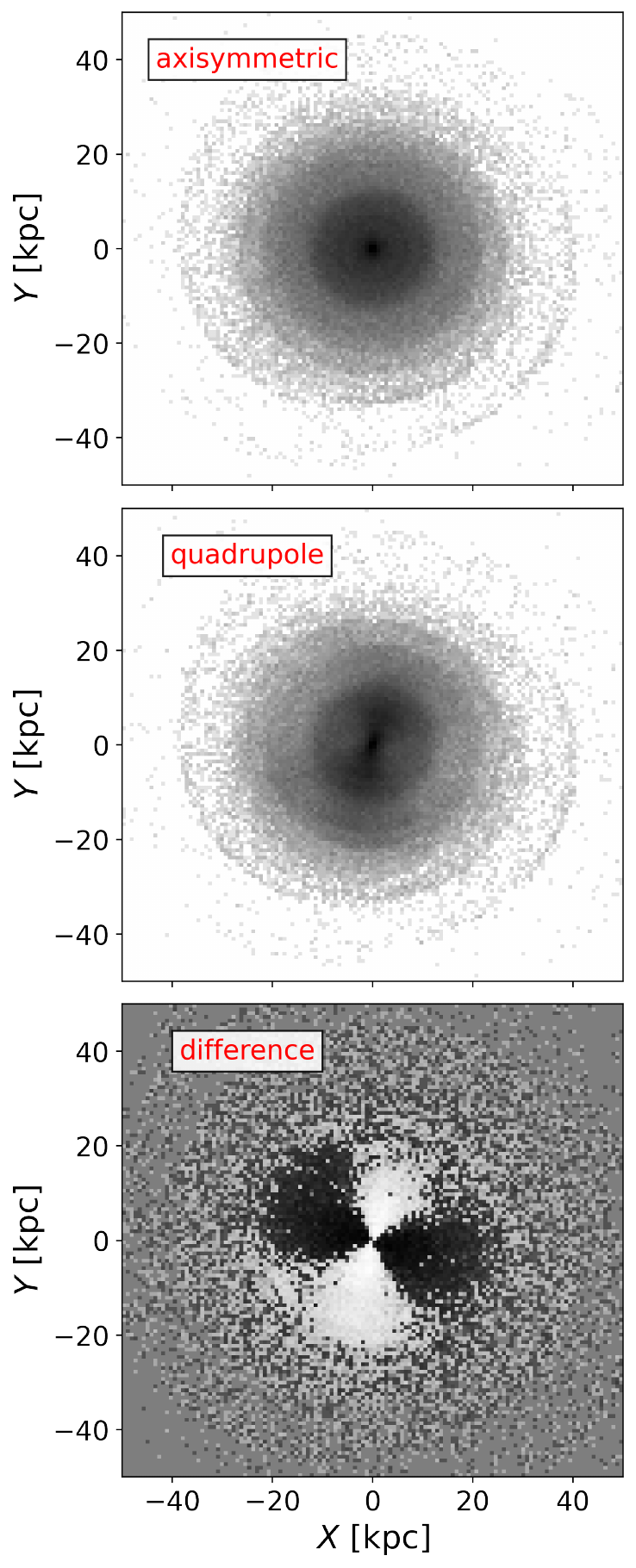}
    \caption{Log density histogram comparison between the $(x,y)$ distribution of the debris after being integrated in the axisymmetric potential and the quadrupole potential for 3 Gyr following the initial 5 Gyr $N$-body simulation. The bottom panel shows the difference between the two distributions.}
    \label{fig:xy_compare}
\end{figure}

In the bottom panels of Fig.~\ref{fig:total_precession}, we show the distribution of $(L_z, \alpha)$ for the selected (low energy, low angular momentum) particles after 8 Gyr of evolution for both the quadrupole and the $N$-body simulation. Compare this with the rightmost panel of Fig.~\ref{fig:alpha_time}, where the orbital planes had precessed substantially more in the axisymmetric case than in either the $N$-body or the quadrupole. This matches up well with Fig.~\ref{fig:compare_lines_quad}, where the striations are completely absent from the low angular momentum regions. Lastly, in Fig.~\ref{fig:xy_compare}, we show how the increased precession rate in the axisymmetric potential causes a washing out of the central elongated structure when viewed from the top down. We plot the $(x,y)$ distribution of particles at the final $t=0$ snapshot after being integrated in the axisymmetric and quadrupole potentials. This effect was similarly explored in \citet[][]{han2022tilt}, and is important for studies on the lifetime of the VOD and HAC.

\begin{figure*}
    \centering
    \includegraphics[width=0.98\textwidth]{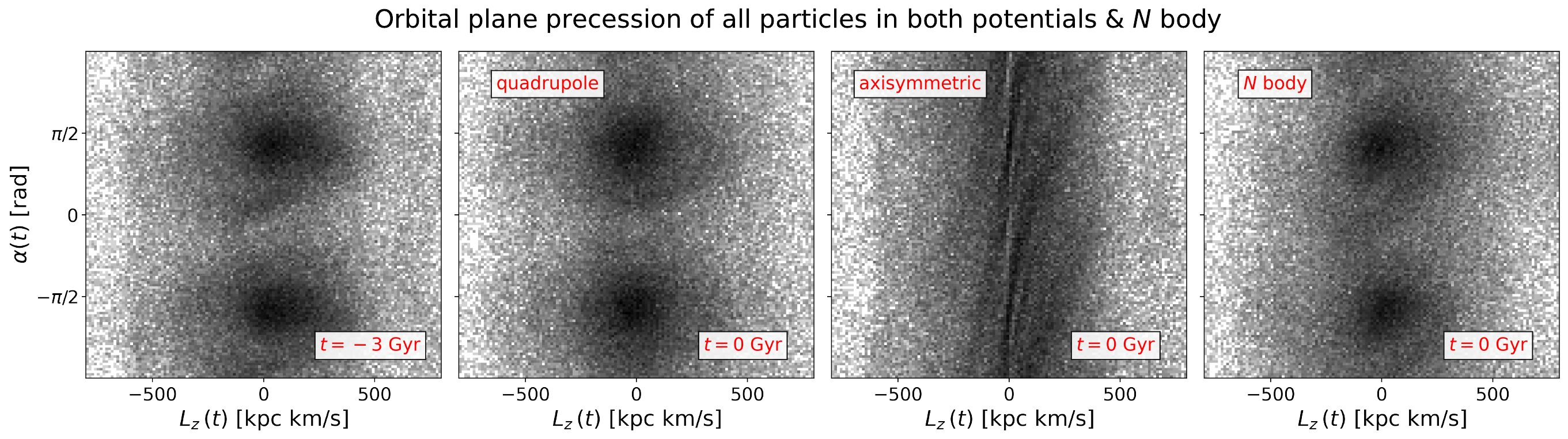}
    \caption{A comparison between the distribution of orbital plane angles $\alpha$ against angular momentum $L_z$ for all particles in each simulation. In the leftmost panel, we show the distribution of particles at the end of the 5 Gyr $N$-body simulation. In the other three panels, we plot the distributions after 3 further Gyrs of evolution for the test particle quadrupole potential, axisymmetric potential, and finally the $N$-body simulation. Note that the quadrupole and $N$-body simulation match up quite closely, whereas the axisymmetric potential causes a dramatic increase in the precession of the orbits.}
    \label{fig:alpha_time_both}
\end{figure*}

\section{Discussion}
This work emphasises the need for accurate representations of the shape of the potential of the Galactic halo. This has been revealed in our simulations in which the LMC interacts with accreted halo debris. Should the MW's halo contain even a slight non-axisymmetry, then the utility of $(L_z, E)$ space for identifying Galactic clusters of a common origin could be thrown into uncertainty, as we can no longer rely on the conservation of angular momentum. Therefore, any groups identified in $(L_z,E)$ space, or equivalent, may be blurred in the $L_z$ direction. We plan to investigate the evolution of known clusters identified in such spaces \citep[e.g.][]{myeong2018discovery, dodd2023gaia} in several potentials of varying degrees of non-axisymmetry to get a measure of how impactful such blurring may be.

The formation of the striations is reliant on our assumed orbit of the LMC, specifically that it is on its first pericentric passage around the MW. However, \citet[][]{vasiliev2023dear} recently explored the possibility that the LMC may in fact have completed prior passages. It seems reasonable to predict that such an orbit could have an ``un-doing'' effect, as the LMC would swing back around the MW and torque the orbits in the opposite direction. It is important to note however, according to the models in \citet[][]{vasiliev2023dear}, if the LMC has undergone previous passages, the last pericentric passage of the LMC likely occured at $r \gtrsim 100$ kpc. This distance would likely to be too great to have any significant impact on the GSE's stellar debris, which is mostly deposited within the central $r \simeq 30$ kpc of the halo. Note also that a previous pericentric passage is estimated to have occured 5 -- 10 Gyr ago. At this time in the MW's history, the GSE is likely to not have fully mixed onto its host, and so much of the stellar halo would not have settled.

An additional further step to explore would be the impact that Sagittarius dwarf spheroidal galaxy (Sgr) may have on $(L_z,E)$ distribution. The nature of Sgr's orbit is similar to the LMC's in that they are both on polar orbits, yet the multiple passages that Sgr has already completed make the situation more complicated. We can make a rough approximation as to the amplitude of the torquing of Sgr, compared to the LMC, by using Eq.~\ref{eq:lim_torque}. If we take $M_{\rm LMC}~=~1.4\times10^{11}$~M$_{\odot}$, $M_{\rm Sgr}~=~4\times10^{8}$~M$_{\odot}$ \citep[][]{vasiliev2020last}, and calculate the torque at each galaxies' respective pericentre of $r_{\rm LMC}~=~50$~kpc and $r_{\rm Sgr}~=~16$~kpc, we get that $\tau_{\rm \, LMC}/\tau_{\rm \, Sgr}~=~11.5$. The effect from Sgr is therefore reduced compared to the LMC, meaning the orbits' $L_z$ will be changed less, and striations may be far less visible. Although, given that Sgr likely had much larger mass at its previous pericentres \citep[e.g.][]{gibbons2017sgr,read2019sgr}, the amplitude of the torquing may not be so different between the two.

Somewhat complementary to both the discussion of alternative LMC orbits and considering the orbit of Sgr, we conducted some additional simulations in which the debris was rotated around the $z$-axis, thus effectively changing the trajectory of the perturbing LMC. These minor experiments confirmed what we already expected -- by adjusting the LMC's orbit to be more directly parallel with the long axis of remnant debris, we reduce the appearance of the striations since the torquing is weaker. By contrast, if the LMC's trajectory is perpendicular to the remnant debris, the torquing is enhanced, and thus the striations become stronger.

\section{Conclusion}\label{sec:summary}

The stellar halo is a component of the Milky Way that provides insight into the formation of the Galaxy. It is populated with a variety of substructures that have different origins. Substructures within the stellar halo are often identified by their kinematic properties, specifically their clustering in integral of motion space. It is therefore vital to understand any external influences on the orbital properties of populations within the stellar halo, in case such an external force imprints on angular momentum substructure. We investigate the effect of a large LMC-like perturber on the stellar halo. Since the GSE likely dominates the stellar halo content, we run an $N$-body simulation of a GSE-MW like merger and investigate the effect that the LMC has on these particles. After sufficient mixing, we take a snapshot of this $N$-body potential and integrate the the orbits as test particles in a two different static multipole potentials: an axisymmetric potential and bisymmetric quadrupole potential. Finally, we allow the LMC to disturb the resulting debris and inspect the changes in $(L_z, E)$ space.

We identify a phenomenon that results from the interaction of our LMC model on our simulated halo particles, when an axisymemtric potential is assumed for the test particles. Specifically, we note that the LMC imprints a series of vertical striations in $(L_z, E)$, resulting from the formation of new overdensities in $L_z$ space, from an original more uniform distribution. The lines form as a result of the following. The value of $L_z$ dictates the precession rate of near radial orbits in the $(x,y)$ plane. This implies that the initial value of $L_z$ dictates the orientation of orbital planes in the $(x,y)$ plane. In the presence of an infalling LMC, the orientation of the orbital planes determines any change in $L_z$, due to the imbalance of the direct force between the LMC and merger debris. This imbalance of forces produces a periodic relationship between orbital plane orientation and $L_z$. This, therefore, implies that there is a periodic relationship between $L_z$ and LMC torque induced change in $L_z$. This periodicity causes bunching along certain values of $L_z$, hence the striations.

Upon instead assuming a quadrupole potential as the static host potential, these vertical striations become far less apparent, especially at low $L_z$. We show that this is due to the increased orbital plane precession rate when axisymmetry is assumed. We note that the quadrupole potential matches up much more closely with a $N$-body simulation ran for a full 8 Gyr. We have shown that even small deviations from axisymmetry can severely influence the precession rate when integrating near-radial orbits, like those accreted from the GSE merger, and lead to the type of biased results that we present in Sec.~\ref{sec:mergerdebris}. While the effect is greatly exaggerated by the enforced axisymmetry in the potential, the torquing of the LMC could still have an effect on the $L_z$ distribution in a similar way that we have shown, given some relationship between orbital orientation and $L_z$. Therefore, these features may be a way of constraining the axisymmetry of the inner halo.

In future work, we hope to more robustly explore how the exact nature of the striations depends on the shape of the host potential for smaller mergers that don't influence the shape of the total potential as much as the GSE does.

\section*{Acknowledgements}

We thank the anonymous referee taking the time to provide insightful comments that improved this work. The authors additionally thank the rest of the Cambridge Streams group, Denis Erkal, Sergey Koposov and Eugene Vasiliev for useful comments. EYD thanks the Science and Technology Facilities Council (STFC) for a PhD studentship (UKRI grant number 2605433). AMD thanks STFC for a PhD studentship (UKRI grant number 2604986).

\section*{Data Availability}

The simulations in this project can be reproduced with publicly available software using the description provided in Sec.~\ref{sec:method}, and the references therein.

\bibliographystyle{mnras}
\bibliography{stripes}

\appendix

\section{Striations in a Smooth Halo}\label{sec:dfdebris}

While we have seen the vertical striations manifest in debris resulting from a merger, which is already heavily substructured, here we look to recreate this effect in a halo formed by sampling a steady state distribution function. We sample a large number particles from a distribution function (described below), implemented in \textsc{Agama} \citep[][]{vasiliev2019agama}, and then make various cuts in $\alpha$ to assess the presence of the vertical striations after interaction with our model of the LMC.  

\subsection{Distribution function}

We adopt the same distribution function for the stellar halo as found in \citet[][]{dillamore2023stellar}, and repeat essentially details here for convenience. Before making any $\alpha$ cuts, we sample $N=5\times10^7$ particles from a double power law halo-like distribution function: 
\begin{align}
    f(\textbf{\textit{J}})&=\frac{M}{(2\pi J_0)^3}\left[1+\left(\frac{J_0}{h(\textbf{\textit{J}})}\right)^\eta\right]^{\Gamma/\eta}\left[1+\left(\frac{h(\textbf{\textit{J}})}{J_0}\right)^\eta\right]^{-B/\eta},\\
    h(\textbf{\textit{J}})&\equiv J_r+|J_\phi|+J_z.
\end{align}
where we take $\Gamma = 2.4$, $B = 4.6$, $J_0 = 3500$ kpc km/s, and steepness $\eta = 10$. This roughly recovers a distribution that resembles the Milky Way halo with inner slope $\gamma \sim 2.5$ and outer slope $\beta \sim 4.5$, and break radius $r \sim 25$ kpc \citep[e.g.][]{watkins2009substructure, deason2011milky}.

\subsection{Results}

\begin{figure*}
    \centering
    \includegraphics[width=0.75\textwidth]{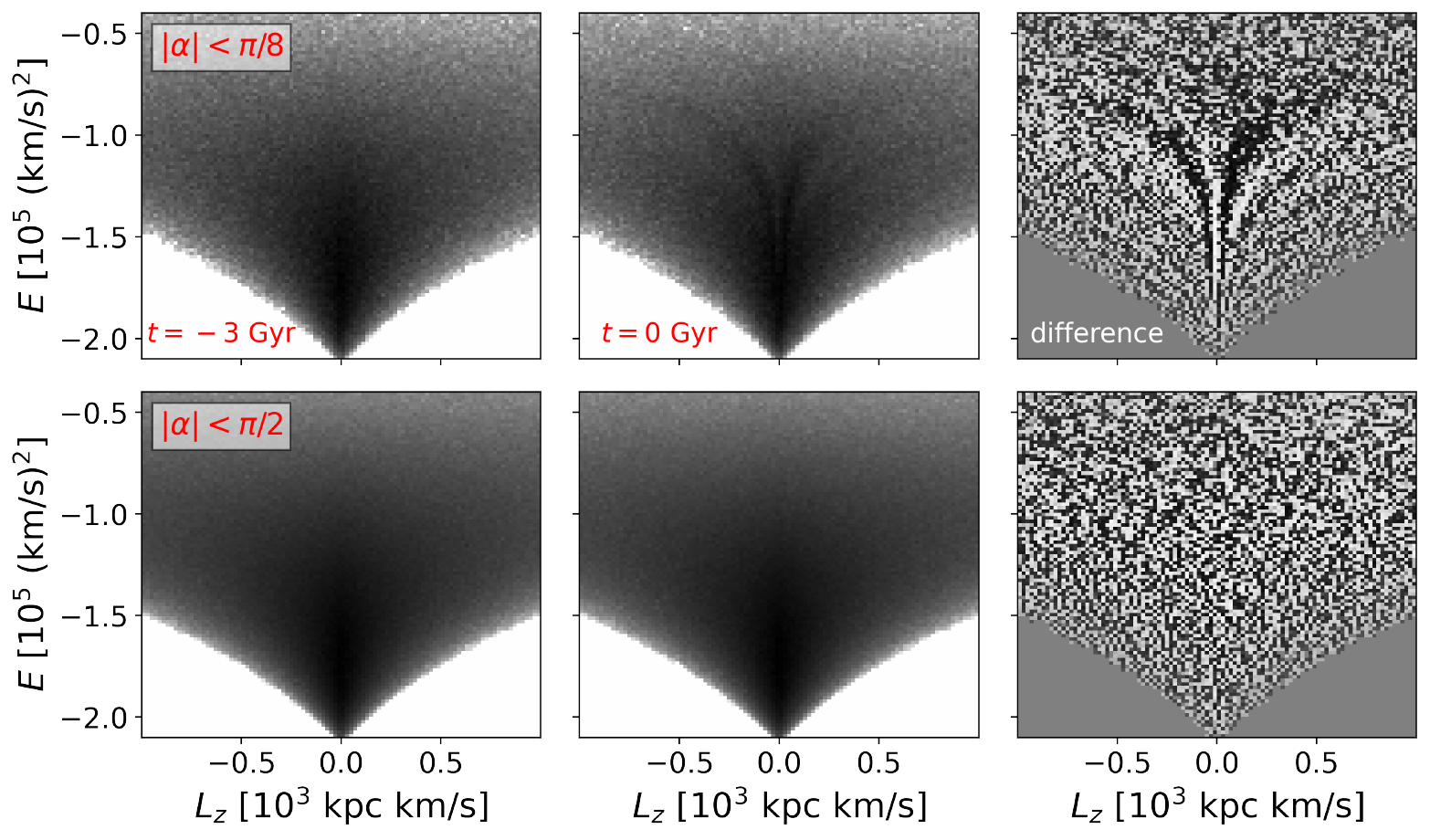}
    \caption{The formation of the striations in a halo formed from a steady state distribution function, in which an orientation angle $\alpha$ cut has been applied. We plot the symmetric log density of $(L_z, E)$. In the top row, a cut of $ -\pi/8 < \alpha < \pi/8$ has been applied, whereas in the bottom row, a much wider cut of $ -\pi/2 < \alpha < \pi/2$ has been applied. The left column shows the $(L_z, E)$ distribution as it was 2 Gyr before the LMC interacts, the middle columns shows it at the final snapshot, with the LMC at $r_{\rm GC}\sim50$ kpc. The right column shows the difference. In the narrow $\alpha$ cut, we see faint vertical lines in the $t=0$ snapshot. By taking the difference, these lines become very clear. However, for the wider distribution, these lines are not present at all.}
    \label{fig:smooth_ELz_alpha}
\end{figure*}

Fig.~\ref{fig:smooth_ELz_alpha} shows the results after letting the LMC infall to the debris for 3 Gyrs. In this example we made two $\alpha$ cuts -- the top row shows a cut with $|\alpha| < \pi/8$, whereas the bottom row shows a cut with $|\alpha| < \pi/2$. In the leftmost column we show the $(L_z, E)$ distributions 3 Gyr before the LMC reaches pericentre, in the middle column we show the distributions when the LMC is at pericentre, and then the rightmost column shows the difference of these two. The narrow cut provides us with $N = 6.2 \times 10^5$ particles, and the wide cut provides us with $N =24.8 \times 10^5$ particles. Despite the 4-fold increase in particle resolution for the wider $\alpha$ cut, we see that there are no visible lines in either the middle of right column of the bottom row. However in the top row, where we have a much more narrow $\alpha$ distribution, there are vertical striations clearly visible. The rightmost column makes this especially clear.

Compared with the simulations in Sec.~\ref{sec:mergerdebris}, the lines are much less clear. This is almost certainly due to the very narrow configuration of the merger debris shown in Fig.~\ref{fig:compare_lines}, which is sparsely populated beyond $|L_z| > 500$ kpc km/s. The halo generated from the distribution function is quite over-saturated, even to values way beyond $|L_z| > 500$ kpc km/s, meaning that the vertical striations get somewhat washed out.

\bsp	
\label{lastpage}
\end{document}